\documentclass[a4paper,UKenglish,cleveref, autoref, thm-restate]{lipics/lipics-v2021}

\makeatletter
\renewcommand\maketitle{\par
  \begingroup
    \thispagestyle{plain}
    \renewcommand\thefootnote{\@fnsymbol\c@footnote}%
    \if@twocolumn
      \ifnum \col@number=\@ne
        \@maketitle
      \else
        \twocolumn[\@maketitle]%
      \fi
    \else
      \newpage
      \global\@topnum\z@   
      \@maketitle
    \fi
    \thispagestyle{plain}\@thanks
  \endgroup
}
\makeatother

\title{On the \ff Algorithm for Online Unit-Interval Coloring} 

\author{Bob {Krekelberg}}{Department of Information and Computing Sciences, Utrecht University, The Netherlands}{b.h.a.f.krekelberg@uu.nl}{https://orcid.org/0009-0000-5517-6095}{}

\author{{Alison Hsiang-Hsuan} {Liu}}{Department of Information and Computing Sciences, Utrecht University, The Netherlands}{h.h.liu@uu.nl}{https://orcid.org/0000-0002-0194-9360}{}

\authorrunning{B. Krekelberg and A.\,H.\,H. Liu} 

\Copyright{Bob Krekelberg and Alison Hsiang-Hsuan Liu} 



\category{}

\relatedversion{} 

\acknowledgements{}

\nolinenumbers 

\usepackage{tikz} 
\usepackage{xspace} 
\usepackage{enumerate} 

\newcommand{\runtitle}[1]{\textbf{#1}}
\newcommand{\orderedInput}{\mathcal{I}}
\newcommand{\result}{\lceil \frac{7}{3}\omega \rceil - 2}
\newcommand{\ff}{\textsc{FirstFit}\xspace} 
\newcommand{\ffc}{\text{\texttt{FF}\xspace}} 
\newcommand{\opt}{\texttt{OPT}\xspace} 
\newcommand{\x}{I} 
\newcommand{\y}{D^\texttt{A}(\x)} 
\newcommand{\z}{D^\texttt{M}(\x)} 
\newcommand{\zbar}{D^\texttt{AM}(\x)} 
\newcommand{\pivot}{I^*} 
\newcommand{\alp}{\alpha} 
\newcommand{\bet}{\beta} 
\newcommand{\gam}{\gamma} 
\newcommand{\delt}{\delta} 
\newcommand{\twins}{\mathcal{T}} 
\newcommand{\nb}{\mathcal{N}} 
\newcommand{\Z}{\mathcal{R}^{\texttt{TDM}}} 
\newcommand{\pivotset}{\mathcal{S}^*} 
\newcommand{\lmr}{\mathcal{A}} 
\newcommand{\nlmr}{\mathcal{M}} 
\newcommand{\rows}[2]{\mathcal{R}_\texttt{#1}(#2)} 
\newcommand{\nrows}[2]{r_\texttt{#1}(#2)} 

\newcommand{\bk}[1]{{\color{Orchid}#1}}
\newcommand{\hide}[1]{}
\newcommand{\hideproof}[1]{}

\newcommand{\fullversion}[1]{{#1}}
\newcommand{\shortversion}[1]{}

\hideLIPIcs
\EventEditors{John Q. Open and Joan R. Access}
\EventNoEds{2}
\EventLongTitle{42nd Conference on Very Important Topics (CVIT 2016)}
\EventShortTitle{CVIT 2016}
\EventAcronym{CVIT}
\EventYear{2016}
\EventDate{December 24--27, 2016}
\EventLocation{Little Whinging, United Kingdom}
\EventLogo{}
\SeriesVolume{42}
\ArticleNo{23}

\begin{document}

\maketitle

\begin{abstract}
    Online interval coloring is a fundamental problem in graph algorithms and scheduling. 
Although the optimal online algorithm for coloring arbitrary-length intervals is known as $3$-competitive (Kierstead and Trotter, 1981), the interest in coloring bounded-length intervals arose recently (Chybowska-Sok\'{o}\l{} et al., 2024). 
On the other hand, people are also interested in the FirstFit algorithm's performance because of its elegance.
The competitive ratio of FirstFit on unit-length intervals is exact $2-\frac{1}{\omega}$, where $\omega$ is the optimal number of colors needed (Epstein and Levy, 2005).
However, for arbitrary-length intervals, the competitive ratio is only known to be between $5$ (Kierstead et al., 2005) and~$8$ (Narayanaswamy and Subhash Babu, 2008). 
It has been open for a long time what the actual competitive ratio of FirstFit is for arbitrary-length intervals.

In this paper, we study the performance of the FirstFit algorithm for the online unit-length intervals coloring problem where the intervals can be either open or closed, which serves a further investigation towards the actual performance of FirstFit.
We develop a sophisticated counting method by generalizing the classic neighborhood bound, which limits the color FirstFit can assign an interval by counting the potential intersections. 
In the generalization, we show that for any interval, there is a critical interval intersecting it that can help reduce the overestimation of the number of intersections, and it further helps bound the color an interval can be assigned. 
The technical challenge then falls on identifying these critical intervals that guarantee the effectiveness of counting. 
Using this new mechanism for bounding the color that FirstFit can assign an interval, we provide a tight analysis of $2\omega$ colors when all intervals have integral endpoints and an upper bound of $\result$ colors for the general case, where $\omega$ is the optimal number of colors needed for the input set of intervals.

\end{abstract}

\section{Introduction}
This work studies the \emph{online open/closed unit-length interval coloring} problem, which is a variant of \emph{online interval coloring} problem.
The input consists of a sequence of unit-length intervals, where each interval is either an \emph{open} interval $(r, r+1) = \{ i \mid r < i < r+1 \}$ or a \emph{closed} interval $[r, r+1] = \{ i \mid r \leq i \leq r+1 \}$.
The intervals are released to the algorithm one by one.
Once an interval is released, the online algorithm has to irrevocably assign the interval a color that is not assigned to any previously released interval that overlaps with the released interval.
The aim is to minimize the colors used.
In this work, we study the performance of the \ff algorithm for coloring the open/closed unit-length intervals.

As a fundamental problem in graph theory and scheduling areas, intensive research is being conducted on the online interval coloring problem.
The optimal online algorithm for the general case where intervals have arbitrary lengths uses at most $3\omega - 2$ colors where $\omega$ is the optimal number of colors~\cite{DBLP:journals/ita/ChrobakS88, kierstead1981extremal, DBLP:conf/mfcs/Slusarek89}.
Recently, interest in coloring bounded-length intervals arose, and better algorithms were found for special families of instances. 
For intervals with the length within some fixed range between $1$ and $\sigma \geq 1$, a $(\sigma + 1)$-competitive algorithm was proposed~\cite{DBLP:journals/ejc/ChybowskaSokol24}, which improved the performance of online algorithms for the cases where $\sigma<2$.

On the other hand, the \ff algorithm, a classical greedy algorithm, is interesting for its elegance. 
The \ff algorithm is currently the best algorithm for (closed) unit-length intervals, which uses $2\omega-1$ colors for any $\omega$-colorable set of unit-length intervals~\cite{DBLP:journals/ita/ChrobakS88, DBLP:conf/icalp/EpsteinL05}.
For arbitrary-length intervals, the competitive ratio is between $5$ and $8$~\cite{DBLP:journals/ejc/KiersteadST16, DBLP:journals/order/NarayanaswamyB08}, and it is open for a long time what the actual competitive ratio is.

The motivation to look into the performance of the \ff algorithm on open/closed unit-length intervals coloring is two-fold.
First, the performance of the \ff algorithm is not fully understood. 
It is known that for unit-length closed intervals, the \ff algorithm is exactly $(2-\frac{1}{\omega})$-competitive, where $\omega$ is the optimal number of colors needed. However, for general instances, the \ff algorithm is at least $5$-competitive.
The performance of the \ff algorithm is unclear for the instances between these cases. 
As the open/closed unit-length intervals case is the smallest possible instance of the coloring problem on bounded length intervals, this case is key to knowing the actual competitive ratio of the \ff algorithm.

Moreover, this research helps us understand the impact of ``$\varepsilon$-uncertainty''.
It can be considered that closed unit-length intervals are $\varepsilon \approx 0$ longer than the open unit-length intervals. 
This difference of $\varepsilon$ increases the complexity of the problem as an open interval can be a proper subset of a closed interval, which was previously not the case when only closed unit-length intervals were considered.
It was shown that the \ff algorithm is at least $2$-competitive~\cite{Curbelo}, meaning that the open/closed unit-length intervals case is strictly ``harder'' for the \ff algorithm than the closed unit-length intervals case. 
Finding the competitive ratio of the \ff algorithm on open/closed unit-length intervals coloring quantifies the power of the uncertainty of $\varepsilon$-difference by understanding how much the adversary can use this uncertainty to trap the \ff algorithm.

\paragraph*{Related work} For the online interval coloring problem, Kierstead and Trotter designed an optimal online algorithm that uses at most $3\omega - 2$ colors on an $\omega$-colorable interval graph~\cite{kierstead1981extremal}.
Independently, Chrobak and \'{S}lusarek found the same upper and lower bounds~\cite{DBLP:journals/ita/ChrobakS88, DBLP:conf/mfcs/Slusarek89}.

Recently, interest in coloring bounded-length intervals arose.
Chybowska-Sok\'{o}\l{} et al.~\cite{DBLP:journals/ejc/ChybowskaSokol24} studied the online interval coloring problem where the size of the intervals is within some fixed range $[1, \sigma_{\geq 1}]$.
When $\sigma = \infty$, it is the previously mentioned general case, and when $\sigma = 1$, it is the unit-length case.
They proposed a $(\sigma + 1)$-competitive algorithm, which improved the performance of the online algorithm for $1<\sigma<2$.
Curbelo~\cite{DBLP:journals/corr/abs-2401-05648} then shows that, in the setting where the interval representation is unknown for the algorithm, for any $\varepsilon > 0$, there exists an $\sigma > 1$, such that any algorithm is at least $(3 - \varepsilon)$-competitive. Matching the upper bound from Kierstead and Trotter for coloring intervals of general length.

\runtitle{\ff.}
On the other hand, despite the aforementioned results that included sophisticated algorithms, people are also interested in the performance of the naive but easy-to-implement \ff{} algorithm.
The exact competitive ratio of the \ff{} algorithm is a long standing open problem.
As early as 1976, Witsenhausen~\cite{DBLP:journals/jct/Witsenhausen76}, and independently, in 1988, Chrobak and \'{S}lusarek~\cite{DBLP:journals/ita/ChrobakS88}, proved that \ff{} is at least 4 competitive.
Later, \'{S}lusarek~\cite{DBLP:conf/mfcs/Slusarek89} improved this bound to 4.45.
The best lower bound known to date is from Kierstead et al.~\cite{DBLP:journals/ejc/KiersteadST16}.
They prove that for every $\varepsilon > 0$, there exist an input such that the \ff{} algorithm uses strictly more than $(5 - \varepsilon)\omega$ colors.

For the upper bounds of the competitive ratio of $\ff$, Kierstead~\cite{DBLP:journals/siamdm/Kierstead88} first proved a constant competitive of $40$ in 1988.
Later, this result was improved to $25.72$ by Kierstead and Qin~\cite{DBLP:journals/dm/KiersteadQ95}.
The next breakthrough was in 2003, when Pemmaraju et al.~\cite{DBLP:journals/talg/PemmarajuRV11} showed that the \ff{} algorithm is at most $10$-competitive.
The current best analysis of $\ff$ was by Narayanaswamy and Subhash Babu~\cite{DBLP:journals/order/NarayanaswamyB08}, who showed that \ff{} uses at most $8\omega - 3$ colors on a set of $\omega$-colorable intervals.

The performance of the \ff{} algorithm has also been of interest in the special cases.
Chybowska-Sok\'{o}\l{} et al.~\cite{DBLP:journals/ejc/ChybowskaSokol24} studied the \ff{} algorithm where the size of the intervals is within some fixed range $[1, \sigma_{\geq 1}]$.
The authors first show a trivial bound of $\lceil \sigma + 1 \rceil\omega$ on the number of colors \ff uses, where $\omega$ is the optimal coloring number.

In the case where only unit-length (closed) intervals are considered, Chrobak and \'{S}lusarek~\cite{DBLP:journals/ita/ChrobakS88}, and independently also Epstein and Levy~\cite{DBLP:conf/icalp/EpsteinL05} proved that the \ff{} algorithm uses exactly $2\omega - 1$ colors for a set of $\omega$-colorable unit-length intervals.

In 2022, instead of only closed intervals as is typically considered, Bir\'{o} and Curbelo proved that \ff{} uses at least $2\omega$ and at most $3\omega - 3$ colors when the optimal number of colors is $\omega$~\cite[Theorem 6.0.1]{Curbelo}.
Furthermore, Curbelo claimed that \ff{} uses at most $3\omega - f(\omega)$ colors for a non-specified function $f$, which increases with $\omega$.
In this paper, we find a better bound for this case.

\runtitle{\ff and counting.}
The analysis of the \ff{} algorithm is generally done via careful counting intersections a critical interval can have.
The elegance of the analysis lies in determining from which interval to count its number of intersecting intervals. 
The first analysis by Kierstead and Qin~\cite{DBLP:journals/siamdm/Kierstead88, DBLP:journals/dm/KiersteadQ95} identify a large set of mutual-intersecting intervals and then carefully count structural properties of the graph, such as the number of overlapping intervals.
The later results by Pemmaraju et al.~\cite{DBLP:journals/talg/PemmarajuRV11} and by Narayanaswamy and Subhash Babu~\cite{DBLP:journals/order/NarayanaswamyB08} first represent the overlapping structure of intervals using ``columns''.
The columns are assigned different labels that encrypt the information of the intervals structure and colors assigned.
Then, the analysis carefully counts the occurrences of each label to bound the total number of colors used by the \ff{} algorithm.
Also, for unit-length (closed) intervals, the analysis carefully counts for each interval at what positions an interval could intersect it and how many of such intervals there could be~\cite{DBLP:journals/ita/ChrobakS88, DBLP:conf/icalp/EpsteinL05}.

\paragraph*{Our contribution}
Following the traditional counting approach, we analyze the $\ff$ algorithm on open/closed unit-length intervals by counting how many intersections an interval can have.

Our main technical contribution is a generalization of the traditional neighborhood bound of the $\ff$ algorithm (Lemma~\ref{lem:prop_ff_equiv}). 
Informally, this lemma shows that the color assigned to interval $\x$ by \ff can be bounded via another interval intersecting $\x$ that is critical in the sense that it is a relatively high-colored interval but with relatively few intersections.
Our results are built on this newly-observed property of \ff. 
In a nutshell, for any interval $\x$, by the fact that the input instance consists only of unit-length open or closed intervals, our analysis identifies the critical interval intersecting $\x$. It shows that the number of intersections is bounded by sophisticated counting. 

For a special case where all intervals have integral endpoints, we prove that $\ff$ is $2$-competitive, which is tight as it matches the $\ff$ algorithm lower bound by Bir\'{o} and Curbelo~\cite{Curbelo}.
Formally, for any instance that can be colored by $\omega$ colors, we show that
\begin{theorem}
    \label{thm:integral}
    For any unit interval graph with open and closed intervals with integral endpoints, the \ff{} algorithm for online coloring uses at most $2\omega$ colors.
\end{theorem}

Then, for unit-length intervals with arbitrary endpoints, we prove the following.
\begin{theorem}
    \label{thm:any}
    For any unit interval graph with open and closed intervals with arbitrary endpoints, the \ff{} algorithm for online coloring uses at most $\result$ colors. 
\end{theorem}

\paragraph*{Paper organization}
Section~\ref{sec:prelim} defines the problem formally. We also introduce the key lemmas, the framework of our analysis, and important concepts that are used heavily throughout the paper.
As a starter, Section~\ref{section:integral_endpoints} shows the proof of $2$-competitiveness (Theorem~\ref{thm:integral}) of \ff on open/closed unit-length intervals with integral endpoints.
We also prove some interesting facts that hold in both the restricted and the more general setting where intervals have arbitrary endpoints.
Section~\ref{section:any_endpoints} then analyze the case where the unit-length intervals have arbitrary endpoints and prove Theorem~\ref{thm:any}.
Finally, we end in Section~\ref{section:conclusion} with concluding remarks.
Due to the page limit, we leave most of the proofs in Section~\ref{section:any_endpoints} to the full version in the appendix.
\label{sec:typesetting-summary}

\section{Preliminaries, key lemmas, and the framework}

The input to our online coloring problem is an ordered set of open and closed unit-length intervals~$\orderedInput = \{I_1, I_2, \cdots\}$, where the ordering of the set of intervals is the order in which they are revealed to the online algorithm. 
That is, $I_j$ is revealed earlier than $I_k$ if $j<k$.
Each interval $I_j\in \orderedInput$ is either an \emph{open} unit-length interval $(r_j, r_j+1)$ or a \emph{closed} unit-length interval $[r_j, r_j+1]$ for some real number $r$.
In either case, we say $r_j$ and $r_j+1$ are the \emph{endpoints} of the interval. 
Note that the value of $r_j$ is unrelated to the order of $I_j$ in $\orderedInput$.
Namely, $r_j$ can be smaller than, equal to, or larger than $r_k$ when $j<k$.
We ignore the suffix of intervals when the reveal ordering is not necessarily to be emphasized.

Two intervals are referred to as \emph{twins} when they are identical in terms of the exact location.
That is, for interval $I = [r, r+1]$, all intervals $I^\prime = [r, r+1]$ are its twins. 
Similarly, for interval $I = (r, r+1)$, all intervals $I^\prime = (r, r+1)$ are its twins.
However, an interval $(r,r+1)$ is not a twin of $[r, r+1]$, and vice versa.
When intervals $I$ and $I^\prime$ are twins, we write $I \equiv I^\prime$.
We further denote the set of twins of $I$ by $\twins(I)$.
Note that we define $I \not \in \twins(I)$ for later usage.

A \emph{proper coloring} of a set of intervals is a function $c : \orderedInput \rightarrow \mathbb{N}$, such that $c(I) \neq c(I^\prime)$ if $I \cap I^\prime \neq \emptyset$.
The algorithm aims to find a proper coloring of the intervals using the least number of colors.

We use terms from graph algorithms. 
The intervals in $\orderedInput$ form an \emph{interval graph}, where each $I\in \orderedInput$ is a vertex in the interval graph, and there is an edge between the vertices corresponding to $I$ and $I^\prime$ if and only if $I \cap I^\prime \neq \emptyset$.
We denote the \emph{neighborhood} of $I$ by $\nb(I)$, which is the set of all the intervals interesting with $I$.

As the coloring depends on intersecting intervals as well as the ordering of the input, we define the $\ell$-\emph{neighborhood} of an interval as its neighborhood at the moment when $I_\ell$ is revealed to the algorithm.
Formally, $\nb_\ell(I_j) = \{ I_k \mid I_j \cap I_k \neq \emptyset \text{ and } k < \ell \}$.\footnote{We only use $\nb_\ell(I_j)$ for $\ell \geq j$.}
Naturally, $\nb(I_j) = \nb_{|\orderedInput|}(I_j) \geq \nb_\ell(I_j)$ for any $\ell \in [j,|\orderedInput|]$.
Note that $\nb_j(I_j)$ is the set of all intervals that intersect with $I_j$ when $I_j$ is revealed.

Interval graphs are \emph{perfect graphs}~\cite{golumbic2004algorithmic}, and the chromatic number of the graph is equal to the maximum clique size $\omega$.
This clique corresponds to the maximum set of mutually intersecting intervals for the corresponding intervals.
We denote $\omega$ as the number of colors an optimal solution uses for the given input $\orderedInput$.

\hide{
\begin{definition}
    \label{def:neighbourhood}
    The $\ell$-\emph{neighborhood} of the interval $I_i \in \orderedInput$ is the set of intervals intersecting with $I_i$, which arrived before $I_\ell$ in the input.
    That is, \\
    $\nb_\ell(I_i) = \{ I_k \mid I_i \cap I_k \neq \emptyset \text{ and } k < \ell \}$.
\end{definition}
}

\medskip

\runtitle{\ff algorithm.}
The \ff algorithm assigns each interval the least available color, that is, using $\ffc(I_j)$ to denote the color assigned to $I_j$ by \ff,
$\ffc(I_j) \leftarrow \min \{ \mathbb{N} \setminus \{ \ffc(I) \mid I \in N_j(I_j) \} \}$.
This greedy strategy provides an upper bound for the color of an interval.
That is, the color of an interval can never be larger than $1$ color above the number of intervals it intersects.
More specifically, the color of the interval will be at most $1$ greater than the size of the neighborhood when it is revealed. 
Formally,
\hide{This gives us the following upper bound for a color assigned by the \ff{} algorithm.}

\begin{lemma}\emph{\textbf{(Neighborhood bound)}}
    \label{lem:prop_ff}
    For any interval $I_j \in \orderedInput$, its color assigned by \ff is at most $1 + |N_j(I_j)|$.
\end{lemma}
\fullversion{
\begin{proof}
    Assume aiming towards contradiction that $\ffc(I_j) \geq 2 + |N_j(I_j)|$.
    Then, according to the pigeonhole principle, there must exist at least one color $1 \leq c < \ffc(I_j)$ that is not assigned to any of the intervals in $N_j(I_j)$.
    Then $\ffc(I_j) = \min \{ \mathbb{N} \setminus \{ \ffc(I) \mid I \in N_j(I_j) \} \} \leq c$, and it leads to a contradiction.
\end{proof}
}

\runtitle{Pivot interval and ideas of further bonding the color.}
In this work, we further generalize Lemma~\ref{lem:prop_ff} by considering a \emph{pivot} interval in the neighborhood of the interval $I$ and use the pivot interval to bound the color $\ffc(I)$ more carefully.
Informally, given any interval~$I$ and a pivot interval $\pivot$ that can be any interval intersecting with $I$, $\ffc(I)$ is no more than~$1$ plus the number of intervals in $I$'s neighborhood with color greater than the color of $\pivot$. 
More formally,

\begin{lemma}\emph{\textbf{(Pivot bound)}}
    \label{lem:prop_ff_equiv}
    For any interval $I_j \in \orderedInput$, let $\pivot \in N(I_j)$ be any interval in $\nb(I_j)$ and let $\pivotset \subset \nb(I_j)$ be the set of intervals in the neighborhood of $I_j$ such that for all $I^\prime\in \pivotset$, $\texttt{FF}(I^\prime)>\texttt{FF}(\pivot)$.
    Then, $\texttt{FF}(I_j) \leq \texttt{FF}(\pivot) + |\pivotset| + 1$.
\end{lemma}

\begin{proof}
    Assume aiming toward contradiction that $\ffc(I_j) \geq 2 + \ffc(\pivot) + |\pivotset|$.
    Then, according to the pigeonhole principle, there must exist at least one color $\ffc(\pivot) < c < \ffc(I_j)$ that is not assigned to any of the intervals in $\pivotset$.
    Then $\ffc(I_j) = \min \{ \mathbb{N} \setminus \{ \ffc(I) \mid I \in N_j(I_j) \} \} \leq c$, and it leads to a contradiction.
\end{proof}

Note that the Neighborhood bound is a special case of the Pivot bound by selecting an empty set as its pivot. In this extreme case, the pivot is colored by $0$ by \ff, and $\pivotset \subseteq \nb(I)$.

Intuitively, for a smaller upper bound of $\ffc(I_j)$, by the Pivot bound, we can choose a~pivot~$\pivot$ with a small color $\ffc(\pivot)$ or choose a~$\pivot$ with a small $|\pivotset|$. 
However, we have a~two-fold challenge. 
First, it is not trivial to find the best choice between minimizing $\ffc(\pivot)$ or minimizing $|\pivotset|$. 
Second, even when a good pivot $\pivot$ is given, it does not necessarily have non-trivial upper bounds of $\ffc(\pivot)$ and $|\pivotset|$.
To cope with the difficulties, we construct a mechanism to balance the two choices of minimizing $\ffc(\pivot)$ or $|\pivotset|$ and different manners to bound $\ffc(\pivot)$ and $|\pivotset|$.
\hide{Ideally, a good selection of pivot interval $\pivot$ with a small $|\pivotset|$ provides a better bound of the color assigned by \ff.
However, if we try to find a pivot $\pivot$ with very small color $\ffc(\pivot)$, it might be the case that $|\pivotset|$ is very large, and vice versa.}

\begin{figure}[t]
    \centering
    \includegraphics[width=0.35\linewidth]{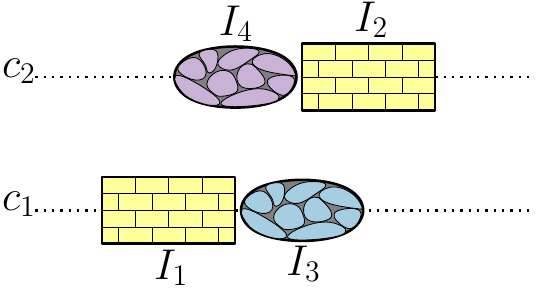}
    \caption{Rows denote the color assigned by \opt, where actual colors denote the color assigned by \ff, where the order is based on the index.}
    \label{fig:example_of_a_row}
\end{figure}

\medskip

\runtitle{Optimal coloring and the image of walls of rocks and bricks.}
In the analysis of \ff algorithm, we compare the solution of the \ff algorithm to an \emph{optimal offline algorithm} \opt that knows the complete input in advance.

We first imagine the intervals in $\orderedInput$ as rocks (open intervals) and bricks (closed intervals) with fixed horizontal positions. 
That is, the rocks and bricks can be shifted vertically but not horizontally. 
We imagine that the optimal solution orients these rocks and bricks into a wall row by row using the fewest rows (without violating the fixed horizontal position constraints).
More specifically, the intervals on the $i$-th row (from the bottom of the wall) are the intervals colored by $i$ according to the optimal solution \opt.
Since \opt uses precisely~$\omega$ colors, the number of rows is exactly $\omega$.
Furthermore, it naturally follows that no two intervals drawn in the same row will intersect with each other, as otherwise \opt admits no proper coloring.

To avoid confusion between the color assigned by the optimal offline algorithm and the color assigned by the \ff algorithm, from now on, we distinguish the color of an interval by \ff or by \opt by its \emph{color} or its \emph{row}. 
More specifically, the color of an interval is the color assigned by \ff algorithm, and the row refers to the color assigned by \opt.
An example of the distinction between colors and rows is shown in Figure~\ref{fig:example_of_a_row}.

\medskip

\runtitle{Relate the intervals and their structure in the wall.}
With the image of rock-and-brick wall in mind, given any interval (which may be a rock or a brick), we partition the rows according to their alignment with the interval $\x$.

\begin{definition}
    For any interval $\x \in \orderedInput$, let $\rows{i}{\x}$ be the set of rows in the optimal solution \opt{}$(\orderedInput)$ that contain $i$ intervals intersecting $\x$.
    And let $\nrows{\texttt{i}}{I}$ denote $|\rows{i}{\x}|$. 
    \hide{, i.e., the number of $\rows{i}{\x}$ rows in the optimal solution \opt{}$(\orderedInput)$ for interval $\x$.}
\end{definition}

Strictly speaking, the set $\rows{i}{\x}$ contains rows instead of containing intervals.
However, by slight abuse of notation, we say that interval $I^\prime \in \rows{i}{\x}$ if $I^\prime$ is in $\nb(\x)$ and colored by one of the colors in  $R_i(\x)$ by the optimal solution.
In other words, $I^\prime \in \rows{i}{\x}$ if $I^\prime$ is an interval overlapping with $\x$ and colored by one of the colors in $\rows{i}{\x}$ by \opt.

Since the intervals are either open unit intervals or closed-unit intervals,
it is clear that not for all $i$ the set $\rows{i}{\x}$ is non-empty.
Formally,
\hide{
Given either an open or a closed unit-length interval $\x$, a row can only be contained in the set $R_i(\x)$ if it is possible that $\x$ intersects with $i$ intervals in that row.
Therefore we shall prove upper bounds on the number of intersections an interval can have per row.
}
\begin{lemma}\label{lem:ris}
    Consider any $I\in \orderedInput$.
    \begin{enumerate}[(a)]
        \item \label{lem:r0_not_empty} $\rows{0}{I}$ contains at least the row where $I$ is, and $\nrows{\texttt{0}}{I} \geq 1$.
        \item \label{lem:closed_intersects_3}
        If $I$ is closed, $\nrows{\texttt{i}}{I} = 0$ for all $\texttt{i}\geq 4$.
        \item \label{lem:open_intersects_2}
        If $I$ is open, $\nrows{\texttt{i}}{I} = 0$ for all $\texttt{i}\geq 3$.
        \item \label{Obs:sumrrr}
        Since \opt uses $\omega$ rows, $\nrows{\texttt{0}}{I} + \nrows{\texttt{1}}{I} + \nrows{\texttt{2}}{I} + \nrows{\texttt{3}}{I} = \omega$.
    \end{enumerate}
\end{lemma}

\hide{
\begin{lemma}
    \label{lem:closed_intersects_3}
    A closed unit-length interval intersects at most $3$ intervals per row.
\end{lemma}
}

\fullversion{
\begin{proof}

\runtitle{(\ref{lem:r0_not_empty})}
    By definition, the rows $\rows{i}{\x}$ correspond to the rows of an optimal solution \opt.
    Therefore, as \opt admits a proper coloring, there is no interval on the same row where $\x$ is contained and intersects $\x$.
    Hence, on the row that contains $\x$, no interval intersects $\x$. Thus, this row is contained in $\rows{0}{\x}$. 
    It follows that $\nrows{0}{\x} \geq 1$.

\runtitle{(\ref{lem:closed_intersects_3})}
    Given a closed unit-length interval $\x = [a, a + 1]$, consider the following 3 points; $a, a + \frac{1}{2}$ and $a + 1$.
    Observe that the distance between these points is strictly smaller than the length of any (open or closed) unit-length interval, and that the endpoints of $\x$ are within these points.
    Then it follows that any interval that intersects interval $\x$ must intersect at least one of the points $a, a + \frac{1}{2}$, or $a + 1$.
    Thus since intervals which are drawn on the same row cannot intersect each other, at most $3$ intervals per row can intersect interval $\x$.

\runtitle{(\ref{lem:open_intersects_2})}
    Given an open unit-length interval $\x = (a, a + 1)$ and an infinitesimal number $\varepsilon$, consider the following 2 points; $a + \varepsilon$ and $a + 1 - \varepsilon$.
    Observe that the distance between these points is strictly smaller than the length of any (open or closed) unit-length interval, and that there exist no points between $a$ and $a + \varepsilon$ and similarly there exist no points between $a + 1$ and $a + 1 - \varepsilon$.
    Then it follows that any interval that intersects interval $\x$ must intersect at least one of the points $a + \varepsilon$ or $a + 1 - \varepsilon$.
    Thus, since intervals that are drawn on the same row cannot intersect each other, at most $2$ intervals per row can intersect interval $\x$.

\runtitle{(\ref{Obs:sumrrr})}
    By definition, the rows in $\rows{i}{\x}$ correspond to the rows of an optimal solution \opt, of which there are $\omega$ such rows.
    We know from (\ref{lem:closed_intersects_3}) that for all $i \geq 4$, $\nrows{i}{\x} = 0$.
    Then, it follows that $\nrows{\texttt{0}}{I} + \nrows{\texttt{1}}{I} + \nrows{\texttt{2}}{I} + \nrows{\texttt{3}}{I} = \omega$.
    Note that this is for any interval $\x$ an exact equality because of the inclusion of $\nrows{0}{\x}$.
\end{proof}
}

By the definition of $\rows{i}{\x}$, the number of intervals that intersect with $\x$ can be upper bounded in terms of $\nrows{i}{\x}$. More specifically,

\begin{observation}
    \label{obs:r_i=N}
    For any interval $\x \in \orderedInput$, $|\nb_\x(\x)| \leq |\nb(\x)| \leq \nrows{\texttt{1}}{\x} + 2\cdot \nrows{\texttt{2}}{\x} + 3\cdot \nrows{\texttt{3}}{\x}$.
\end{observation}

Recall from Lemma~\ref{lem:ris} (\ref{lem:closed_intersects_3}) and (\ref{lem:open_intersects_2}) that only closed intervals $I$ can have non-empty $\rows{3}{I}$. 
Further, there is only a single way for a closed interval to intersect three intervals in the same row. 
Therefore, all rows in $\rows{3}{I}$ contain three sets of intervals, where the intervals in each set are twins.
Formally, given any interval $I = [r, r+1]$ where $r$ is some real number, we partition the intervals in $\nb(I)$ into three sets:
\begin{itemize}
    \item $\twins(I)$: Intervals $I^\prime \equiv [r, r+1]$.
    
    \item $\lmr(I)$: Intervals that are twins of intervals in $\rows{3}{I}$. More specifically, $\lmr(I)$ consists of the intervals $I^\prime \equiv [r-1, r]$, $I^\prime \equiv (r, r+1)$, and $I^\prime \equiv [r+1, r+2]$.
    
    \item $\nlmr(I)$: All other intervals in $\nb(I)$.
\end{itemize}

\label{sec:prelim}

\section*{Framework}
\label{section:general}
Given that the optimal coloring uses $\omega$ colors on $\orderedInput$. 
For any interval $\x \in \orderedInput$, we bound its color from \ff by using the Neighborhood bound (Lemma~\ref{lem:prop_ff}) and the Pivot bound (Lemma~\ref{lem:prop_ff_equiv}).
More specifically, we pick a pivot $\pivot$ with special structural property. 
Using the property, we bound the size of the corresponding $\pivotset$. 
On the other hand, we bound the color $\ffc(\pivot)$ using the Neighborhood bound. 
Finally, we use the Pivot bound to bound the color of $\x$ using the pivot $\pivot$.

Practically, we first identify the ``easy cases'' where the $\ffc(\x)$ is at most~$2\omega$ as desired.
Then, after peeling off these easy cases, we focus on the tough kernel of the analysis where there is no trivial choice for $\pivot$ such that the sum of $\ffc(\pivot)$ and $|\pivotset|$ is small.
For this tough kernel of analysis, we identify two mediocre choices for $\pivot$ for which the sum of $\ffc(\pivot)$ and $|\pivotset|$ cannot both be large at the same time.

\medskip

\runtitle{Intervals $\x$ with small $\nrows{3}{\x}$.}
These intervals have relatively small neighborhoods. Formally, \shortversion{}{by the Neighborhood bound (Lemma~\ref{lem:prop_ff}) and Lemma~\ref{lem:ris} (\ref{Obs:sumrrr}), we can show the following lemma.}
\begin{lemma}
    \label{lem:integral_open_and_closed_assumtion_not_hold}
    For each interval $\x \in \orderedInput$, 
    if $\nrows{3}{\x} < \nrows{1}{\x} + 2$, then $\ffc(\x) \leq 2\omega$.
\end{lemma}

\fullversion{
\begin{proof}
    By the Neighborhood bound (Lemma~\ref{lem:prop_ff}), $\ffc(\x) \leq 1 + |\nb_{\x}(\x)|\leq 1 + |\nb(\x)|$. 
    By Lemma~\ref{lem:ris} (\ref{Obs:sumrrr}), $\ffc(\x)\leq 1 + \nrows{1}{\x} + 2\nrows{2}{\x} + 3\nrows{3}{\x}$.
    The given condition $\nrows{3}{\x} < \nrows{1}{\x} + 2$ implies that $\ffc(\x) \leq 2 + 2\nrows{1}{\x} + 2\nrows{2}{\x} + 2\nrows{3}{\x}$. 
    Since $\nrows{0}{\x} \geq 1$, it follows that $\nrows{1}{\x} + \nrows{2}{\x} + \nrows{3}{\x} \leq \omega - 1$.
    Thus, $\ffc(\x) \leq 2\omega$.  
\end{proof}
}

Since any interval in $\rows{3}{\x}$ must be in $\lmr(\x)$, an empty $\lmr(\x)$ implies $\nrows{3}{\x} = 0$. 
Hence, we have a useful corollary:

\begin{corollary}
    \label{cor:y_exists}
    If $\lmr(\x) = \emptyset$, then $\ffc(\x) \leq 2\omega$.
\end{corollary}

\medskip

\runtitle{Open intervals.\footnote{This is a generalization from the results of Chrobak and {\'{S}}lusarek~\cite{DBLP:journals/ita/ChrobakS88} and Epstein and Levy~\cite{DBLP:conf/icalp/EpsteinL05}, which showed for closed unit-length intervals, \ff uses exactly $2\omega - 1$ colors.}}
By Lemma~\ref{lem:ris} (\ref{lem:open_intersects_2}), open intervals cannot have too many intersections per row. More specifically,

\begin{theorem}
    \label{lem:open_interval_small_intersection}
    For each open interval $\x \in \orderedInput$, $\ffc(\x) \leq 2\omega - 1$.
\end{theorem}

\fullversion{
\begin{proof}
    By Lemma~\ref{lem:ris} (\ref{lem:open_intersects_2}), an open interval $\x$ intersects at most $2$ intervals per row.
    Furthermore, since there are $\omega$ rows and $\x \in \rows{0}{\x}$, at most $\omega - 1$ rows remain to have any intervals intersecting $\x$.
    It follows that $|\nb_\x(\x)| \leq |\nb(\x)| \leq 2\omega - 2$.
    By the Neighborhood bound, $\ffc(\x) \leq 2\omega - 1$.
\end{proof}
}

\runtitle{Closed intervals $\x$ with large $\nrows{3}{\x}$.}
Intuitively, because the intervals in $\twins(\x)$ have an identical neighborhood to interval $\x$, these intervals are tricky to deal with.
This leaves us to consider the intervals in the sets $\lmr(\x)$ and $\nlmr(\x)$ as potential pivot intervals.
In the ideal scenario, by picking as a pivot the interval with the largest color in the sets $\lmr(\x)$ and $\nlmr(\x)$, $\pivotset \subseteq \twins(\x)$.
Using the Neighborhood bound, we bound the color of this largest interval.
However, it is not always possible to pick such a pivot such that $\pivotset \subseteq \twins(\x)$.
In the case that $\pivotset \not\subseteq \twins(\x)$, we look more closely at what additional intervals are part of this set.
Finally, we apply the Pivot bound in order to prove an upper bound on the color of interval $\x$.
\hide{
First, we use the Neighborhood bound to bound the colors of intervals in $\lmr(\x)$. In \bk{Lemma~\ref{lem:integral_uvw}}, we show that as these intervals have an aligned structure, they have relatively small neighborhoods.
Once we bound the color of an interval in $\lmr(\x)$, it can be used as a pivot in the Pivot bound.

\bk{Second, we look into intervals in the set $\nlmr(\x)$.}
The intervals in $\nlmr(\x)$ have misaligned positions with $\x$ but can have higher colors than the intervals in $\lmr(\x)$ and cause a higher color of $\x$.
\bk{Therefore, they can also be considered as a pivot.
However, it requires some extra effort to bound the colors of intervals in $\nlmr(\x)$.
In Lemma~\ref{lem:any_z}, we show that if some structural properties are met, the color of intervals in $\nlmr(\x)$ is small.}
However, \bk{in Lemma~\ref{lem:any_z}}, we show that the size of $\nlmr(\x)$ is small (otherwise, we get the desired bound directly). 
Then, by using intervals in $\lmr(\x)$ as a pivot, we show that $\pivotset \subseteq \nlmr(\x)$ and bound the color of $\x$ by the Pivot bound.
}
\medskip

In Sections~\ref{section:integral_endpoints} and~\ref{section:any_endpoints}, we will focus on the last case where $\x$ is a closed interval with large~$\nrows{3}{\x}$.

\section{Warm-up: Regular case where endpoints are integral}
\label{section:integral_endpoints}
As a starter, we consider a special case where all intervals have integral endpoints. 
More formally, for any interval $\x \in \orderedInput$, $I = (i, i+1)$ for some integer $i$ if $I$ is open, and $I = [i, i+1]$ for some integer $i$ if $I$ is closed.
In this section, we show that any $\x \in \orderedInput$ has color $\ffc(\x)\leq 2\omega$, where $\omega$ is the number of colors used by the optimal solution.
Note that in the following, we focus on the closed intervals $\x \in \orderedInput$ with $\nrows{3}{\x} \geq \nrows{1}{\x}+2$ (by Lemma~\ref{lem:open_interval_small_intersection} and Lemma~\ref{lem:integral_open_and_closed_assumtion_not_hold}).

\begin{figure}[t]
\begin{minipage}{0.4\textwidth}
\centering
\includegraphics[scale=0.4]{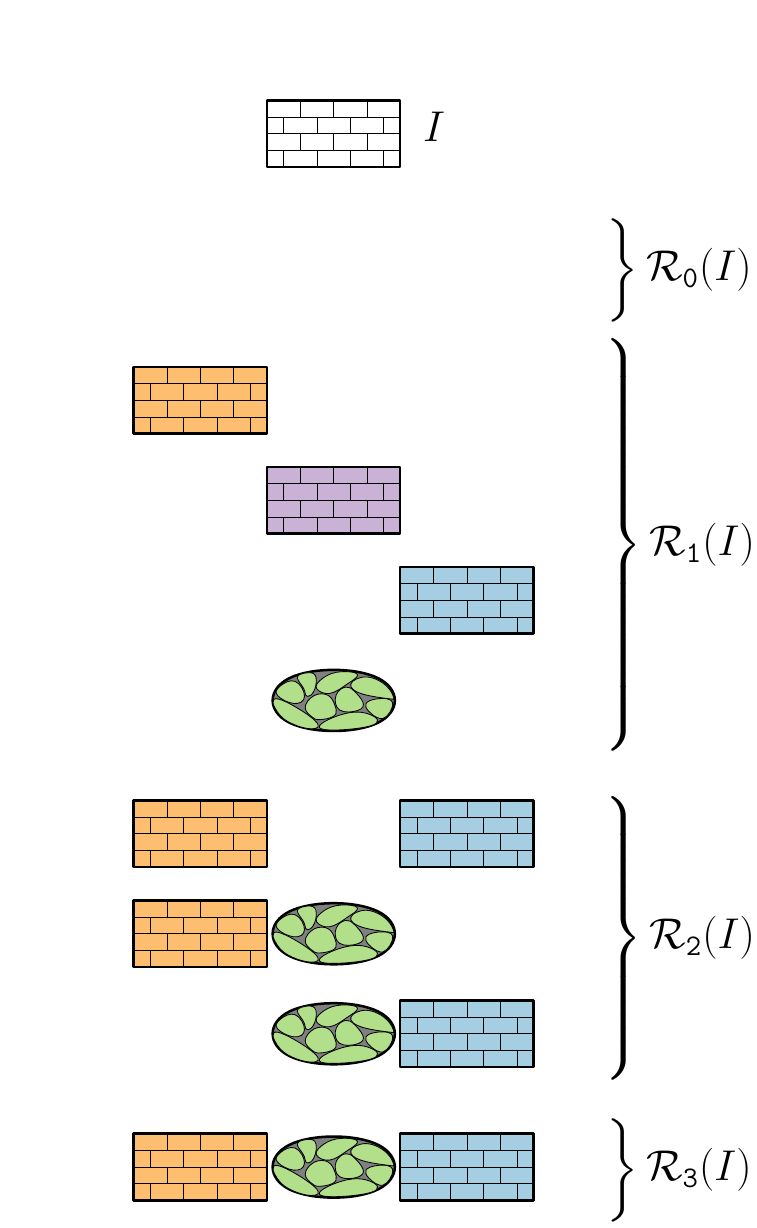}
\caption{
All possible positions of intervals intersecting $\x$ per type of row.
The intervals in $\lmr(\x)$ are drawn in orange, green and blue, the intervals in $\twins(\x)$ in purple.
}
\label{fig:column_structure_in_rows}
\end{minipage}%
\hspace{0.2\textwidth}%
\begin{minipage}{0.4\textwidth}
\centering
\includegraphics[scale=0.4]{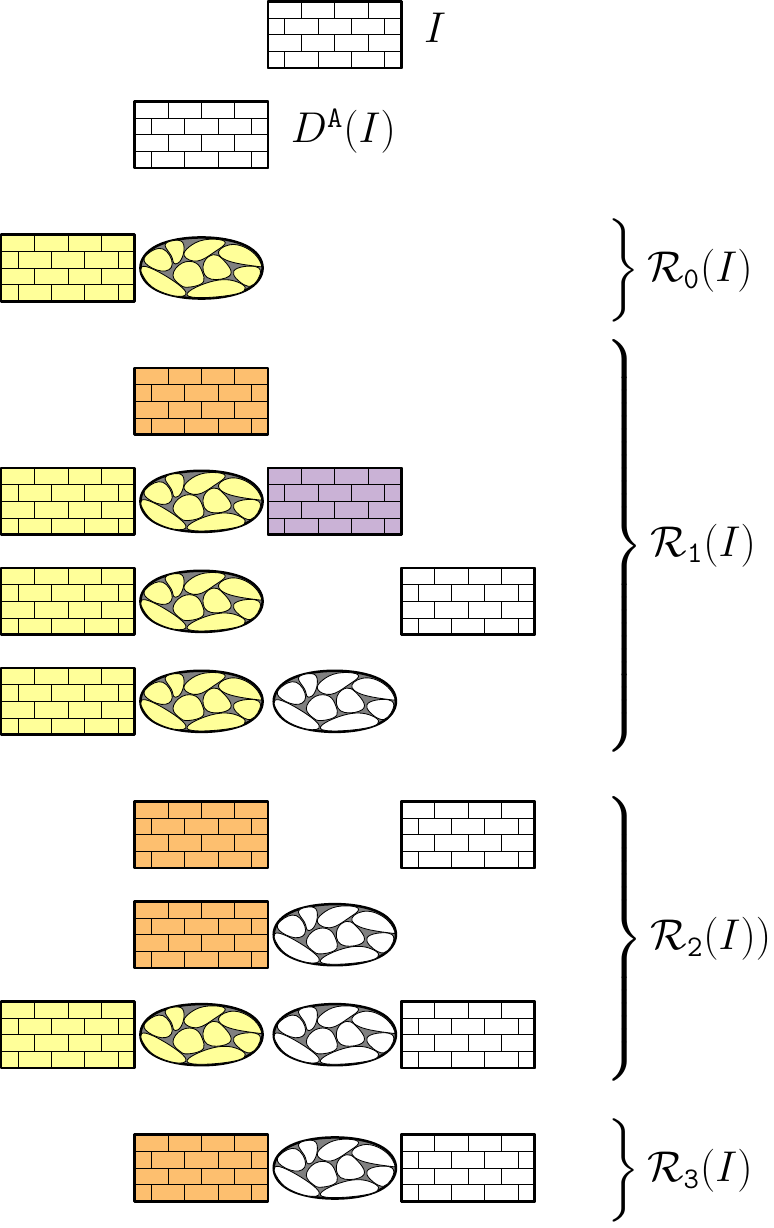}
\caption{
All possible intersections with interval $\y$ per type of row.
The intervals from $\lmr(\x)$ that intersect with $\y$ are drawn in orange, the intervals in $\twins(\x)$ in purple and the remaining intervals intersecting $\y$ are drawn in yellow.}
\label{fig:intersections_with_v}
\end{minipage}
\end{figure}

For such a set of integral-endpoints intervals, it is an important property that $\nlmr(\x) = \emptyset$ for any $\x \in \orderedInput$.
That is, the intervals in $\nb(\x)$ are either in $\twins(\x)$ (twins of $\x$) or in $\lmr(\x)$ (twins of the intervals in $\rows{3}{\x}$).
An example is shown in Figure~\ref{fig:column_structure_in_rows}.

\paragraph*{Pick a pivot $\pivot$}

For any closed intervals $\x \in \orderedInput$ with $\nrows{3}{\x} \geq \nrows{1}{\x}+2$, we first identify the interval in $\lmr(\x)$ with the highest color by \ff. 

\begin{definition}\emph{\textbf{(Dominating interval $\y$ in $\lmr(\x)$)}}
    \label{def:y}
    Let interval $\y \in \lmr(\x)$ be the highest colored interval. That is, there is no $I^\prime \in \lmr(\x)$ such that $\ffc(I^\prime) > \ffc(\y)$.
\end{definition}

Note that $\y$ must exist as $|\lmr(\x)| \geq \nrows{3}{\x} \geq \nrows{1}{\x} + 2 \geq 2$.

In this analysis, we want to bound $\ffc(\x)$ by using $\y$ as a pivot $\pivot$ and applying the Pivot bound (Lemma~\ref{lem:prop_ff_equiv}).

\paragraph*{Bound the size of $\pivotset$}
Given that we use $\y$ as the pivot $\pivot$, we bound the size of $\pivotset$.
Recall that $\pivotset$ is the set of intervals in $\nb(\x)$ which are assigned higher color than $\ffc(\pivot)$.
Since all intervals in $\orderedInput$ have integral endpoints, any interval in $\nb(\x)$ is either in $\lmr(\x)$ or in $\twins(\x)$.
Therefore, by the definition of $\y$, only intervals in $\twins(\x)$ are possible to contribute to $\pivotset$.
Moreover, since we focus on $\x$ that is closed, all intervals in $\twins(\x)$ must be in the rows in $\rows{1}{\x}$. 
We use the following definition to represent the fraction of rows in $\rows{1}{\x}$ that cannot contribute to $\pivotset$.

\begin{definition}
    \label{def:alpha}
    Given any $\hat{I} \in \rows{1}{\x}$, $\alp(\hat{I}) \in [0,1]$ denotes the fraction of rows $\mathcal{R} \in \rows{1}{\x}$ where the interval $I^\prime$ in $\mathcal{R} \cap \nb(\x)$ satisfies \emph{1)} $I^\prime \notin \lmr(\x)$ and \emph{2)} $\ffc(I^\prime) \leq \ffc(\hat{I})$.
\end{definition}

Note that we make Definition~\ref{def:alpha} more general for later usage. 
For the integral-endpoints case, it is sufficient to set $\hat{I}$ as $\y$, and $\alp(\y) \cdot \nrows{1}{\x}$ is the number of intervals in $\twins(\x)$ that are not in $\pivotset$.

\begin{lemma}
    \label{lem:integral_S}
    If all intervals in $\orderedInput$ have integral endpoints, by selecting $\y$ as a pivot~$\pivot$, $|\pivotset| = (1-\alp(\y)) \cdot \nrows{1}{\x}$.
\end{lemma}

\begin{proof}
    First, by the definition of $\rows{0}{\x}$, $\x$ does not intersect any interval in $\rows{0}{\x}$. Thus, no interval in $\rows{0}{\x}$ can be in $\pivotset$.
    Next, as all intervals have integral endpoints, all intervals in $\rows{2}{\x}$, $\rows{3}{\x}$, or $\rows{1}{\x} \setminus \twins(\x)$ are in the sets $\lmr(\x)$.
    Thus, all these intervals are assigned a color below $\ffc(\y)$ and cannot contribute to $\pivotset$.
    The remaining intervals are those in $\twins(\x)$.
    By definition, $\alp(\y) \cdot \nrows{1}{\x}$ of these intervals have a color at most $\ffc(\y)$.
    Hence, the $(1 - \alp(\y)) \cdot \nrows{1}{\x}$ intervals are the only intervals intersecting $\x$ that have a color greater than $\ffc(\y)$.
    That is, $|\pivotset| = (1 - \alp(\y)) \cdot \nrows{1}{\x}$.
\end{proof}

\paragraph*{Bound the color $\ffc(\pivot)$}

We bound the color of $\y$ using the Neighborhood bound (Lemma~\ref{lem:prop_ff}).
In general, apart from intervals in $\twins(\x)$, the number of intervals any interval in $\lmr(\x)$ can intersect is shown in the following observation (also see Figure~\ref{fig:intersections_with_v}):

\begin{observation}
    \label{obs:integral}
    Any interval in the set $\lmr(\x)$ intersects
    \begin{enumerate}[(a)]
        \item at most $2$ intervals per row in $\rows{0}{\x}$,
        \item at most $2$ intervals which are not twins with $\x$ per row in $\rows{1}{\x}$,
        \item at most $2$ intervals per row in $\rows{2}{\x}$, and
        \item exactly $1$ interval per row in $\rows{3}{\x}$.
    \end{enumerate}
\end{observation}

According to the definition of $\y$, Observation~\ref{obs:integral} applies to $\y$.

\begin{lemma}
    \label{lem:integral_uvw}
    The color of interval $\y$, $\ffc(\y) \leq 2 \omega + \alp(\y) \cdot \nrows{1}{\x} - \nrows{3}{\x}$.
\end{lemma}

\begin{proof}
    By the Neighborhood bound, $\ffc(\y) \leq 1+|N(\y)|$.
    By Observation~\ref{obs:integral}, there are at most $2\nrows{0}{\x} + 2\nrows{1}{\x} + 2\nrows{2}{\x} + \nrows{3}{\x}$ intervals in $\lmr(\x)$ that are in $\nb(\y)$. 
    Together with the $\alp(\y) \cdot \nrows{1}{\x}$ intervals that ``occupied'' the smaller colors from the choice of $\y$ at the moment when $\y$ arrives, $|\nb_{\y}(\y)| \leq 2\nrows{0}{\x} + 2\nrows{1}{\x} + \alp(\y) \cdot \nrows{1}{\x} + 2\nrows{2}{\x} + \nrows{3}{\x}$.
    Observe that by this bound, we count the row that contains interval $\y$ itself. 
    Since $\y$ cannot intersect any interval on this row, we should subtract at least $1$ from this bound.
    Thus,
    \begin{align*}
        \ffc(\y) &\leq 1 + 2\nrows{0}{\x} + 2\nrows{1}{\x} + \alp(\y)\cdot \nrows{1}{\x} + 2\nrows{2}{\x} + \nrows{3}{\x} - 1\\
        &= 2 \omega + \alp(\y)\cdot \nrows{1}{\x} - \nrows{3}{\x}
    \end{align*}
\end{proof}

Note that due to the abstraction of $\alp(\y)$, this proof does not rely on the property of integral endpoints.
Therefore, Observation~\ref{obs:integral} also holds for the general case where the intervals in $\orderedInput$ have arbitrary endpoints.

\begin{corollary}
    \label{cor:integral_LMR}
    Given instance $\orderedInput$ with arbitrary endpoints, the color of interval $\y$ is at most $2\omega + \alp(\y) \cdot \nrows{1}{\x} - \nrows{3}{\x}$.
\end{corollary}

\paragraph*{Proof of Theorem~\ref{thm:integral}}
Now, we are ready to prove Theorem~\ref{thm:integral} that for any $\orderedInput$ with open or closed unit-length intervals that have integral endpoints, $\ffc(\x) \leq 2\omega$ for all $\x \in \orderedInput$, where $\omega$ is the optimal color needed for properly coloring $\orderedInput$.

\begin{proof}
    By Lemma~\ref{lem:integral_open_and_closed_assumtion_not_hold}, we assume that $\nrows{1}{\x} \leq \nrows{3}{\x}-2$. 
    Moreover, by Lemma~\ref{lem:open_interval_small_intersection}, we focus on closed $\x$.
    According to Corollary~\ref{cor:y_exists}, we assume that there exists a dominating interval $\y$ in $\lmr(\x)$ that has the highest color by \ff.
    By Lemma~\ref{lem:integral_uvw}, $\ffc(\y) \leq 2 \omega + \alp(\y) \cdot \nrows{1}{\x} - \nrows{3}{\x}$.

    We now use the Pivot bound to bound the color of interval $\x$ from above, using $\y$ as the role of $\pivot$.
    By Lemma~\ref{lem:integral_S}, $|\pivotset| = (1 - \alp(\y)) \cdot \nrows{1}{\x}$.
    \shortversion{Altogether, we have $\ffc(\x) \leq \ffc(\pivot) + |\pivotset| + 1 \leq 2 \omega - 1$}
    \fullversion{
    It follows that
    \begin{align*}
        \ffc(\x) &\leq \ffc(\pivot) + |\pivotset| + 1 \\
        &= \ffc(\y) + (1 - \alp(\y)) \cdot \nrows{1}{\x} + 1\\
        &\leq 2\omega + \alp(\y)\cdot \nrows{1}{\x} - \nrows{3}{\x} + (1 - \alp(\y)) \cdot \nrows{1}{\x} + 1\\
        &\leq 2\omega + \nrows{1}{\x} - \nrows{3}{\x} + 1 \\
        &\leq 2 \omega - 1
    \end{align*}
    }
\end{proof}

\section{General case: Intervals with arbitrary endpoints}
\label{section:any_endpoints}

In this section, we consider the instance $\orderedInput$, where intervals have arbitrary endpoints. 
More specifically, by Lemma~\ref{lem:integral_open_and_closed_assumtion_not_hold} and Lemma~\ref{lem:open_interval_small_intersection}, we focus on $\x \in \orderedInput$ that are closed and $\nrows{3}{\x} \geq \nrows{1}{\x} + 2$.
Recall from Section~\ref{section:integral_endpoints} that we partition $\nb(\x)$ into $\twins(\x)$, $\lmr(\x)$, and $\nlmr(\x)$, that is, the set of twins, the set of intervals that are twins of intervals in $\rows{3}{\x}$, and the rest of intervals with endpoints ``misaligned'' with the endpoints of $\x$.

\shortversion{Primarily, we pick a pivot interval $\pivot$ and bound its color using the Neighborhood bound. 
The choice of pivot should provide a plausible way to bound the size of the corresponding~$\pivotset$.
However, applying the Pivot bound to bound the color $\ffc(\x)$ in this general case is more challenging than the integral-endpoints case due to the misaligned structure of intervals.
Directly applying the same technique as in the intergral-endpoints case results in a ratio of~$3\omega$ (see the full version).}
\fullversion{
Applying the Pivot bound to bound the color $\ffc(\x)$ in this general case is more challenging than the integral-endpoints case. 
First, unlike in the integral-endpoints case where all intervals that intersect $\x$ must be either in $\twins(\x)$ or in $\lmr(\x)$, the candidates in $\pivotset$ can also be in $\nlmr(\x)$. 
These intervals in $\nlmr(\x)$ can cross-interact with other intervals, making it difficult to bound the size of $\pivotset$.
Second, selecting a good pivot interval $\pivot$ in the general case is more challenging. }
\fullversion{We first show that using the same strategy as we used for the integral-endpoints case only guarantees a bound of $\ffc(\x) \leq 3\omega$.

\paragraph*{A naive attempt}

As in the integral-endpoints case, let $\pivot$ be $\y$, the interval with the highest color in~$\lmr(\x)$. 
In the general case, any interval in the sets $\rows{1}{\x}$ and $\rows{2}{\x}$ can be in $\nlmr(\x)$ and can therefore be a candidate of $\pivotset$.
If we now apply the same strategy as in the integral-endpoints setting, it is no longer the case that the set $\pivotset$ is a subset of $\rows{1}{\x}$.
Then, in each row $\mathcal{R} \in \rows{2}{\x}$, it can be the case that two intervals (of which one intersects $\x$ and the other does not) are in the neighborhood of the pivot $\pivot$.
Furthermore, the remaining interval on row $\mathcal{R}$ that also intersects with $\x$ can be considered for the set $\pivotset$, resulting in us counting $3$ intervals per row in $\rows{2}{\x}$.
Thus, the bound tends to $3\omega$ when $\nrows{1}{\x}$ and $\nrows{3}{\x}$ are both small.
To work around this, we need to closely look at the color of the intervals in $\rows{2}{\x}$. 
}

\fullversion{
\subsection{Roadmap}
Primarily, we pick a pivot interval $\pivot$ and bound its color using the Neighborhood bound. 
The choice of pivot should provide a plausible way to bound the size of the corresponding $\pivotset$. }

In the general case, we have another possible candidate of $\pivot$.
Symmetric to the dominating aligned interval $\y$, we define interval $\z$ as the dominating interval in rows in $\rows{2}{\x}$ that is misaligned with $\x$ and assigned the highest color by $\ff$.
Formally,

\begin{definition}
    \shortversion{\label{def:z_short}}
    \fullversion{\label{def:z}}
    \emph{\textbf{(Dominating interval $\z$ in $\rows{2}{\x}\cap\nlmr(\x)$)}}
    Let interval $\z$ be the interval in $\rows{2}{\x}\cap\nlmr(\x)$ with the highest color by $\ff$. That is, for all $I^\prime \in \rows{2}{\x}$, $\ffc(I^\prime) \leq \ffc(\z)$.
\end{definition}

A special property of the selection of $\z$ is that since $\z \in \nlmr(\x)$, any row in $\rows{0}{\x}, \rows{1}{\x}$ or $\rows{3}{\x}$, cannot be in $\rows{3}{\z}$\fullversion{ (Lemma~\ref{lem:R3z_in_R2x_short})}.
Therefore, all rows in $\rows{3}{\z}$ must also be in $\rows{2}{\x}$.

\shortversion{
\begin{lemma}
    \label{lem:R3z_in_R2x_short}
    $\rows{3}{\z} \subseteq \rows{2}{\x}$.
    \end{lemma}
}

In the general case, we pick one of $\y$ and $\z$ as the pivot and use the Pivot bound to bound the color $\ffc(\x)$.
More specifically, we consider the following two cases.


\shortversion{
\paragraph*{Case (1): $\ffc(\z) \leq \ffc(\y)$}
}
\fullversion{
\subparagraph*{(1)}
$\ffc(\z) \leq \ffc(\y)$.}

In this case, the misaligned intervals in $\rows{2}{\x}$ are not in $\pivotset$ since they all have colors of at most $\ffc(\z) \leq \ffc(\y)$.
Using $\y$ as the pivot and following a strategy\shortversion{: }
\fullversion{similar to the case of integral-endpoints, we can show that $\ffc(\x) \leq 2\omega$ (Lemma~\ref{lem:c_u>=c_z}).}

\shortversion{

\smallskip

\begin{lemma}
    \label{lem:c_u>=c_z_short}
    For any interval $\x \in \orderedInput$, if $\ffc(\y) \geq \ffc(\z)$, then $\ffc(\x) \leq 2\omega$.
\end{lemma}
}

\shortversion{\paragraph*{Case (2): $\ffc(\z) > \ffc(\y)$}}
\fullversion{\subparagraph*{(2)}
$\ffc(\z) > \ffc(\y)$.}

In this case, applying the Pivot bound with $\pivot = \y$ leads to a bound of $\ffc(\x) \leq 3 \omega$ as shown in the naive attempt.
To be precise, it is no longer true that $\pivotset \subseteq \rows{1}{\x}$, and any interval in $\rows{2}{\x}$ with a color greater than $\ffc(\y)$ is also a candidate of $\pivotset$.

An alternative strategy is to pick the interval $\z$ as the pivot in this case.
However, in the general case, the intervals in $\rows{1}{\z}$ are not particularly structured,
More precisely, we do not know which intervals also intersect with $\x$, and the neighborhood of $\z$ can be huge. 
When~$\z$ has a large number of intersections, using the Neighborhood bound to bound the color $\ffc(\z)$ can lead to a high bound and further lead to a high bound of $\ffc(\x)$ by the Pivot bound.
To deal with this situation, we first identify the rows containing intervals in $\rows{1}{\z}$ that we know their locations, namely, the intervals that are twins with $\z$.

\smallskip

\begin{definition}
    \shortversion{\label{def:Z_short}}
    \fullversion{\label{def:Z}}
    Let the set of rows in $\rows{2}{\x}$ with intervals identical to interval $\z$ be 
    \[ \Z (\x ) = \{ \text{row } \mathcal{R} \in \rows{2}{\x} \mid \mathcal{R} \text{ contains at least one interval that is in } \twins(\z)  \}.\]
\end{definition}

\shortversion{
    In combination with the set $\Z(\x)$ we are interested in the set $\rows{3}{\z}$.       
    Similar to the definition of $\alp$, we define fractions of specific $\rows{2}{\x}$ rows, that it, $\gam$ and $\delt$:
    
    \begin{definition}
    \label{def:gamma_delta_short}
    Given $\z \in \rows{2}{\x}$, 
        \begin{enumerate}[(a)]
            \item $\bet \in [0,1]$ denotes the fraction of rows $\mathcal{R} \in \rows{2}{\x}$ where the interval $I^\prime$ in $\mathcal{R} \cap \nb(\x)$ is in $\Z(\x)$, and
            \item $\gam \in [0,1]$ denotes the fraction of rows $\mathcal{R} \in \rows{2}{\x}$ where the interval $I^\prime$ in $\mathcal{R} \cap \nb(\x)$ is in $\rows{3}{\z}$.
        \end{enumerate}
    \end{definition}
}


Note that $\Z(\x) \subseteq \rows{1}{\z}$, and thus $\nrows{3}{\z} \leq |\Z(\x)|$ implies that $\nrows{3}{\z} \leq \nrows{1}{\z}$. On the contrary, $\z$ potentially has a large neighborhood if $\nrows{3}{\z} > |\Z(\x)|$.
According to the size of $|\Z(\x)|$, we consider the following cases \textbf{2.a} and \textbf{2.b}.

\shortversion{\paragraph*{Case (2.a): $\nrows{3}{\z} \leq |\Z(\x)|$}}
\fullversion{\subparagraph*{(2.a)}
$\nrows{3}{\z} \leq |\Z(\x)|$.}

In this case, we pick $\z$ as the pivot $\pivot$.
By the selection of $\z$ and $\ffc(\z) > \ffc(\y)$, the only intervals that can contribute to $\pivotset$ are those in the set $\rows{1}{\x}$.
Furthermore, by $|\Z(\x)| \geq \nrows{3}{\z}$, the neighborhood of $\z$ is small, which enables us to use the Neighborhood bound to prove an upper bound the color of interval $\z$\shortversion{:} \fullversion{(Lemma~\ref{lem:any_z}).}
\shortversion{

\smallskip

\begin{lemma}
    \label{lem:any_z_short}
    The color of interval $\z$ is at most $c(\z) \leq \omega + \alp(\z) \cdot \nrows{1}{\x} - \bet  \cdot \nrows{2}{\x} + \gam  \cdot \nrows{2}{\x}  + \nrows{2}{\x} + \nrows{3}{\x}$. 
\end{lemma}

By selecting $\z$ as the pivot $\pivot$, the following bound of $|\pivotset|$ can be shown by the Neighborhood bound together with bounding the intersection of intervals in $\nlmr(\x)$.
\begin{lemma}
    \label{lem:pivotset_z_short}
    When $\ffc(\z) > \ffc(\y)$, by selecting $\z$ as pivot $\pivot$, $ |\pivotset| = (1 - \alp(\z)) \cdot \nrows{1}{\x}$.
\end{lemma}
}
Then, by the Pivot bound\shortversion{:} \fullversion{($\ffc(\x)$ is bounded by $2\omega$ Lemma~\ref{lem:gamma<delta}).}
\shortversion{
\begin{lemma}
    \shortversion{\label{lem:gamma<delta_short}}
    If $|\Z(\x)| \geq \nrows{3}{\z}$, then $\ffc(\x) \leq 2\omega$.
\end{lemma}
}

\shortversion{
\paragraph*{Case (2.b): $\nrows{3}{\z} > |\Z(\x)|$}
}
\fullversion{\subparagraph*{(2.b)}
$\nrows{3}{\z} > |\Z(\x)|$.}

This is the most technical part of our analysis.
In this case, $\z$ may intersect more than~$2$ intervals per row on average and have a large neighborhood, and using the Neighborhood bound on $\ffc(\z)$ can lead to a large bound.
We apply our framework recursively on $\z$ by first showing that the intervals in $\rows{3}{\z}$ rows do not intersect many intervals when $\nrows{3}{\z}$ is large\fullversion{ (Lemma~\ref{lem:any_z_bar})}.
Specifically, we let $\zbar$ be the highest colored interval in $\rows{3}{\z}$ that intersects with interval $x$.

\smallskip

\begin{definition}
    \shortversion{\label{def:z_bar_short}}
    \fullversion{\label{def:z_bar}}
    Let interval $\zbar \in \rows{3}{\z}$ be the dominating interval that has the highest color such that there exists no $I^\prime \in \rows{3}{\z}$ that intersects $\x$ such that $\ffc(I^\prime) > \ffc(\zbar)$.\footnote{Recall that we say an interval $I^\prime$ is in $\rows{i}{\x}$ if it is in $\nb(\x)$ and is colored by one of the colors in $\rows{i}{\x}$ by the optimal solution.}
\end{definition}

It is critical to use $\y$, $\z$, or $\zbar$ as the pivot according to whether $\ffc(\zbar)$ is larger than $\ffc(\y)$.

\shortversion{
\paragraph*{Case (2.b.ii): $\ffc(\zbar) \geq \ffc(\y)$}
}
\fullversion{
\subparagraph*{(2.b.i)}
$\ffc(\zbar) \geq \ffc(\y)$.}

In this case, we apply the Pivot bound by selecting $\zbar$ as the pivot $\pivot$.
Recall that~$\y$ is the dominating interval in $\rows{3}{\x}$.
\shortversion{
Observe that by the definition of interval $\z$, $\ffc(\zbar) < \ffc(\z)$.
Hence, analogous to parameter $\alp$, we define $\delt$.

\smallskip

\begin{definition}
    \label{def:delta_short}
    Given any interval $\hat{I}$, $\delt(\hat{I}) \in [0, 1]$ denotes the fraction of rows $\mathcal{R} \in \rows{2}{\x} \setminus \rows{3}{\z}$ where interval $I^\prime$ in $\mathcal{R} \cap \nb(\x) \cap \nb(\z)$ satisfies $\ffc(I^\prime) \leq \ffc(\hat{I})$.
\end{definition}
}
The condition $\ffc(\zbar) \geq \ffc(\y)$ and the selection of $\zbar$ ensure that no interval in $\rows{3}{\x}$ can be in $\pivotset$.
Furthermore, by the definition of $\zbar$, no interval in $\rows{3}{\z}$ can contribute to $\pivotset$.
Therefore, $\pivotset$ is a subset of intervals in $\rows{1}{\x}$ and $\rows{2}{\x}\setminus \rows{3}{\z}$.

\shortversion{
\begin{lemma}
    \label{lem:pivotset_zbar_short}
    By selecting $\zbar$ as pivot $\pivot$, $|\pivotset| = (1-\alp(\zbar)) \cdot \nrows{1}{\x} + (1 - \delt(\zbar) \cdot (1 - \gam) \cdot 2\nrows{2}{\x}$.
\end{lemma}
}

The definition of $\rows{3}{\z}$ and the selection of $\zbar$ guarantee that a row in the set $\rows{3}{\zbar}$ must contain an interval identical to the interval $\z$ \fullversion{(see Figure~\ref{fig:y_intersections})}.
Thus, $\rows{3}{\zbar} \subseteq \Z(\x)$, and $\nrows{3}{\zbar} \leq |\Z(\x)|$.
Moreover, since $\twins(\zbar) \subseteq \rows{1}{\zbar}$, and every $\rows{3}{\z}$ contains a twin of $\zbar$, $\nrows{1}{\zbar} \geq \nrows{3}{\z}$. 
Together with the condition $\nrows{3}{\z} > |\Z(\x)|$, it guarantees that 
$\nrows{1}{\zbar} \geq \nrows{3}{\z} > |\Z(\x)| \geq \nrows{3}{\zbar}$.
Therefore, $\zbar$ intersects at most $2$ intervals per row on average, and $\ffc(\zbar)$ can be bounded by the Neighborhood bound effectively.
\shortversion{
\begin{lemma}
    \label{lem:any_z_bar_short}
    The color of interval $\zbar$, $\ffc(\zbar) \leq \nrows{0}{\x} + (1 + \alp(\zbar)) \cdot \nrows{1}{\x} + \delt(\zbar) \cdot (1 - \gam ) \cdot 2\nrows{2}{\x} + \bet  \cdot \nrows{2}{\x} + \gam  \cdot \nrows{2}{\x} + 2\nrows{3}{\x}$.
\end{lemma}
}
\fullversion{With the bound $\pivotset \subseteq \rows{1}{\x} \cup (\rows{2}{\x} \setminus \rows{3}{\z})$, the upper bound $\ffc(\x) \leq 2\omega$ (Lemma~\ref{lem:cy>=cu}).}
\shortversion{
Then, by the Pivot bound:
\begin{lemma}
    \label{lem:cy>=cu_short}
    If $\ffc(\zbar) \geq \ffc(\y)$, then $\ffc(\x) \leq 2\omega$   
\end{lemma}
}

\shortversion{
\paragraph*{Case (2.b.ii): $\ffc(\zbar) < \ffc(\y)$}
}
\fullversion{
\subparagraph*{(2.b.ii)}
$\ffc(\zbar) < \ffc(\y)$.}


This case is the tough kernel of the analysis.
In this case, considering using $\y$, $\z$, or $\zbar$ as the pivot only is not sufficient to have an upper bound of $\ffc(\x)$ that is smaller than $3\omega$.
Recall that this case is a subcase of case $\ffc(\y) < \ffc(\z)$, picking $\y$ as the pivot does not stop any interval in $\rows{2}{\x}$ from contributing to $\pivotset$. 
On the other hand, as $\nrows{3}{\z} > |\Z(\x)|$, $\nrows{3}{\z}$ may be much larger than $\nrows{1}{\z}$.
Applying the Neighborhood bound yields a huge upper bound of $\ffc(\pivot)$.
Finally, condition $\ffc(\zbar) < \ffc(\y)$ implies that when $\zbar$ is selected as the pivot, $\pivotset$ could contain any interval in $\rows{3}{\x}$ and become too large.

To deal with this tough case, we express the upper bounds obtained by using $\y$ as the pivot and using $\z$ as the pivot in terms of $|\Z(\x)|$ and $\nrows{2}{\z}$\fullversion{ (Lemmas~\ref{lem:any_z} and~\ref{lem:any_uvw})}. 
These formulations show that the upper bounds from applying the Pivot bound using interval $\y$ as $\pivot$ and using interval $\z$ as $\pivot$ form a trade-off.
More specifically, the upper bound using $\y$ is maximized when $|\Z(\x)| = \emptyset$, while upper bound using $\z$ is maximized when $|\Z(\x)| = \rows{2}{\x}$.
Then, we use that in any circumstance, the upper bound is at most $2\omega + \frac{|\Z(\x)|}{2}$, which is strictly less than $2\omega + \frac{\nrows{2}{\x}}{2}\cdot\omega$ 
(Definition~\ref{def:Z}). 
\fullversion{Since $\nrows{2}{\x}$ can be upper bounded by $\frac{2}{3}\omega - 1$ (Corollary~\ref{cor:r2_bound}), the color assigned to interval $\x$ is bounded by $\result$ (Theorem~\ref{thm:any}).}
\fullversion{
\subsection{Analysis}

\paragraph*{Case (1): $\ffc(\y) \geq \ffc(\z)$}

We start by proving that in case (1), that is, when $\ffc(\y) \geq \ffc(\z)$, the color of interval $\x$ can be bounded by $2\omega$.

\medskip

\runtitle{Pick a pivot $\pivot$}.
    Given the condition where interval $\y$ has a color larger or equal to interval $\z$, picking $\y$ as a pivot provides a smaller bound of $|\pivotset|$.
    Therefore, we pick $\y$ as the pivot $\pivot$.

\medskip

\runtitle{Bound the color of $\ffc(\pivot)$.}
    Next, we bound on the color of the pivot $\ffc(\pivot)$.
    By Corollary~\ref{cor:integral_LMR}, $\ffc(\pivot) = \ffc(\y) \leq 2\omega + \alp (\y) \cdot \nrows{1}{\x} - \nrows{3}{\x}$, where $\alp(\y)$ is the fraction of rows in $\rows{1}{\x}$ that contains intervals in $\twins(\x) \cup \nlmr(\x)$ and having color not higher than $\y$.

\medskip

\runtitle{Bound the size of $\pivotset$}.
Given that we use $\y$ as pivot $\pivot$, we bound the size of $\pivotset$.
Recall that $\pivotset$ is the set of intervals in $\nb(\x)$ which are assigned colors strictly higher than $\ffc(\pivot)$.
Since $\ffc(\y) \geq \ffc(\z)$, only intervals in $\twins(\x)$ are possible to contribute to~$\pivotset$.

\begin{lemma}
    \label{lem:pivotset_y}
    When $\ffc(\y) \geq \ffc(\z)$, by selecting $\y$ as pivot $\pivot$, $|\pivotset| = (1 - \alp(\y)) \cdot \nrows{1}{\x}$.
\end{lemma}

\begin{proof}
    First, by the definition of $\rows{0}{\x}$, $\x$ does not intersect any interval in $\rows{0}{\x}$.
    Thus, no interval in $\rows{0}{\x}$ can be in $\pivotset$.
    Next, the intervals in $\rows{2}{\x}$ are either in $\lmr(\x)$ or in $\nlmr(\x)$.
    Moreover, since $\ffc(\y) \geq \ffc(\z)$, all intervals in $\rows{2}{\x}$ have a color below $\ffc(\y)$ and cannot contribute to $\pivotset$.
    Similarly, as all intervals in $\rows{3}{\x}$ are in $\lmr(\x)$, all intervals in $\rows{3}{\x}$ have a color below $\ffc(\y)$ and cannot contribute to $\pivotset$.
    The remaining intervals are the intervals in $\rows{1}{\x}$.
    By definition, $\alp(\y) \cdot \nrows{1}{\x}$ of these intervals have a color at most $\ffc(\y)$.
    Hence, the remaining $(1 - \alp(\y) ) \cdot \nrows{1}{\x}$ are the only intervals intersecting $\x$ that have a color greater than $\ffc(\y)$. That is, $|\pivotset| = (1 - \alp(\y)) \cdot \nrows{1}{\x}$.
\end{proof}

\medskip

\runtitle{Bound the color of $\ffc(\x)$.}
Now, we wrap up the analysis of the case where $\ffc(\y) \geq \ffc(\z)$ using the bounds of $\ffc(\pivot)$ and $|\pivotset|$.

\begin{lemma}
    \fullversion{\label{lem:c_u>=c_z}}
    For any interval $\x \in \orderedInput$, if $\ffc(\y) \geq \ffc(\z)$, then $\ffc(\x) \leq 2\omega$.
\end{lemma}

\begin{proof}
    By Lemma~\ref{lem:integral_open_and_closed_assumtion_not_hold} and Lemma~\ref{lem:open_interval_small_intersection}, $\ffc(\x) \leq 2\omega$ if $\x$ is open or $\nrows{3}{\x} < \nrows{1}{\x} + 2$.
    Thus, in the following, we focus on closed intervals $\x$ with $\nrows{1}{\x} \leq \nrows{3}{\x} - 2$.
    
    By taking $\y$ as pivot $\pivot$, according to Corollary~\ref{cor:integral_LMR}, $\ffc(\y) \leq 2\omega + \alp(\y) \cdot \nrows{1}{\x} - \nrows{3}{\x}$.
    On the other hand, by  Lemma~\ref{lem:pivotset_y}, $|\pivotset| = (1 - \alp(\y)) \cdot \nrows{1}{\x}$.
    Thus,
    \begin{align*}
        \ffc(\x) &\leq \ffc(\pivot) + |\pivotset| + 1\\
        &\leq \ffc(\y) + (1 - \alp(\y) ) \cdot \nrows{1}{\x} + 1\\
        &\leq 2\omega + \nrows{1}{\x} - \nrows{3}{\x} + 1\\
        &\leq 2\omega - 1
    \end{align*}
\end{proof}

\hide{
In the case where interval $\z$ is in $\lmr(\x)$, then, from the definition of $\y$, it must be the case that $\ffc(\y) \geq \ffc(\z)$.
Therefore, the next result follows directly from Lemma~\ref{lem:c_u>=c_z}.

\begin{corollary}
    \label{cor:z_in_pos_u_v}
    If $\z \in \lmr(\x)$, then $\ffc(\x) \leq 2\omega$.
\end{corollary}
}

\medskip

\paragraph*{Case (2): $\ffc(\y) < \ffc(\z)$}

From now on, we focus on the case where $\ffc(\y) < \ffc(\z)$.

\medskip

\runtitle{Pick a pivot $\pivot$.}
    The condition where $\ffc(\y) < \ffc(\z)$ guarantees that all intervals in the set $\lmr(\x)$ have colors at most $\ffc(\z)$.
    Therefore, for the sake of the size of the $\pivotset$, it is more efficient to pick the interval $\z$ as a pivot than picking the interval~$\y$.
    Therefore, we pick interval $\z$ as pivot $\pivot$.


Recall from Definition~\ref{def:z} that $\z$ is the interval with the highest color in the rows in~$\rows{2}{\x}$. 
The selection of $\z$ attains a good structural property on $\rows{3}{\z}$ with regards to $\rows{2}{\x}$:

\begin{lemma}
    \label{lem:R3z_in_R2x}
    $\rows{3}{\z} \subseteq \rows{2}{\x}$.
\end{lemma}

\begin{proof}
    By case distinction on $\rows{i}{\x}$, we prove by contradiction that there is no intersection between $\rows{3}{\z} \cap \rows{i}{\x}$.
    
    \textbf{(1)} Suppose on the contrary that $\rows{3}{\z} \cap \rows{0}{\x} \neq \emptyset$.
    If there exists a row with $3$ intervals that intersect $\z$, but none of them intersects $\x$, then the intervals $\x$ and $\z$ must not intersect.
    Thus, $\z \notin \nlmr(\x)$, which is a contradiction by the definition of $\z$.

    \medskip
    
    \textbf{(2)} Suppose on the contrary that $\rows{3}{\z} \cap \rows{1}{\x} \neq \emptyset$.
    If there exists a row with $3$ intervals that intersect $\z$, but only one of those intervals intersects $\x$. 
    Then, either the interval~$\z$ is in the set $\lmr(\x)$ or $\x$ and $\z$ do not intersect.
    In either case, $\z \not \in \nlmr(\x)$, which is a contradiction by the definition of $\z$.
    
    \medskip
    
    \textbf{(3)} Suppose on the contrary that $\rows{3}{\z} \cap \rows{3}{\x} \neq \emptyset$. 
    If there exists a row with $3$ intervals that intersect both $\x$ and $\z$, then interval $\z$ must be in the set $\twins(\x)$.
    Then, $\z \not \in \nlmr(\x)$, which is a contradiction by the definition of $\z$.
    
    \medskip
    
    By \textbf{(1)} to \textbf{(3)}, $\rows{3}{\z} \cap \left( \rows{0}{\x} \cup \rows{1}{\x} \cup \rows{3}{\x} \right) = \emptyset$.
    Therefore, it follows that $\rows{3}{\z} \subseteq \rows{2}{\x}$.
\end{proof}

Recall from Definition~\ref{def:Z} that the set $\Z(\x) \subset \rows{2}{\x}$ contains all intervals in an~$\rows{2}{\x}$ row that are twins of interval $\z$. 
\hide{And the set $\rows{3}{\z}$ contains all rows in which~$3$ intervals intersect interval $\z$.
This set $\rows{3}{\z}$ must also be a subset of $\rows{2}{\x}$.}
Furthermore, recall from Definition~\ref{def:z_bar} that interval $\zbar$ is the highest colored interval in $\rows{3}{\z}$ that intersects with $\x$.
Observe that the intersection of the sets $\Z(\x)$ and $\rows{3}{\z}$ must be empty, as a row cannot have both $3$ intervals intersecting $\z$ and an interval identical to $\z$, and therefore $|\Z(\x)| + \nrows{3}{\z} \leq \nrows{2}{\x}$.

\begin{figure}[t]
\begin{minipage}{0.35\textwidth}
\centering
\includegraphics[width=\textwidth]{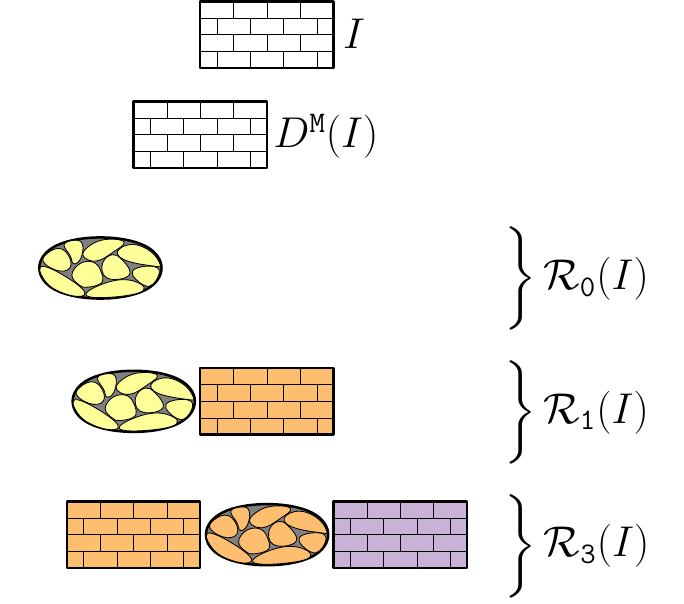}
\end{minipage}%
\hspace{0.1\textwidth}%
\begin{minipage}{0.55\textwidth}
\centering
\includegraphics[width=\textwidth]{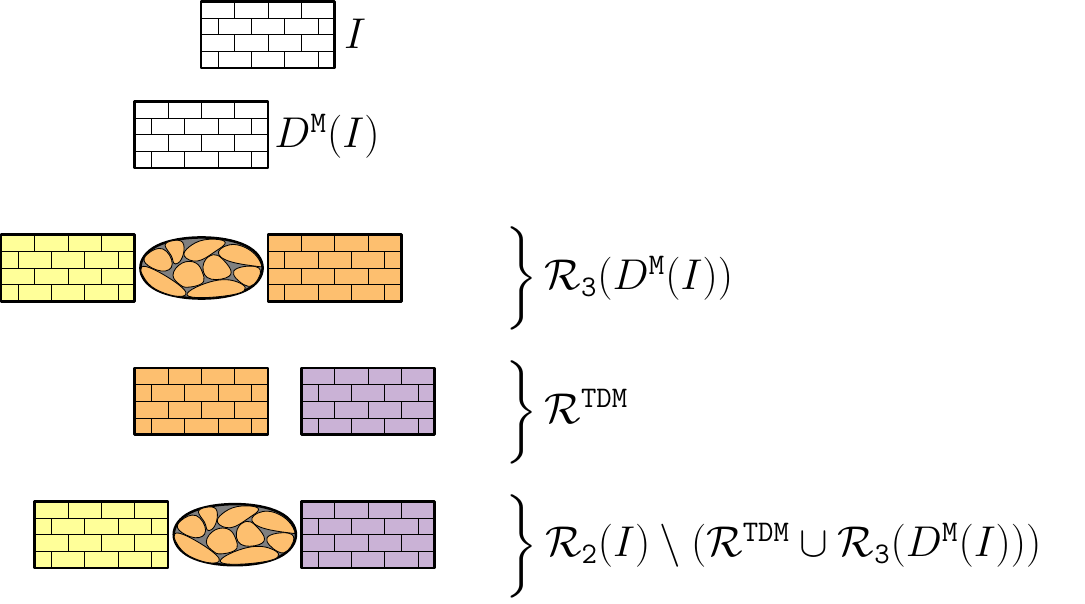}
\end{minipage}
\caption{
An exemplary non-exhaustive list of possible rows per type of row.
Yellow intervals intersect only interval $\z$, purple intervals intersect only $\x$ and orange intervals intersect both $\z$ and $\x$.
}
\label{fig:r2_and_z_intersections}
\end{figure}

\medskip

\runtitle{Bound the color of $\ffc(\pivot)$.}
We first provide observations on the maximum number of intersections intervals in $\nlmr(\x)$ and specifically $\z$ have in each type of row (also see Figure~\ref{fig:r2_and_z_intersections}).

\begin{observation}
    \label{obs:any_r2}
    Any interval in the set $\nlmr(\x)$ intersects
    \begin{enumerate}[(a)]
        \item at most $1$ interval, other than $\x$, per row in $\rows{0}{\x}$,
        \item at most $1$ interval, which does not intersect interval $\x$, per row in $\rows{1}{\x}$, and
        \item exactly $2$ intervals per row in $\rows{3}{\x}$.
    \end{enumerate}
\end{observation}

\begin{observation}
    \label{obs:any_z}
    Interval $\z$ intersects
    \begin{enumerate}[(a)]
        \item exactly $3$ intervals per row in $\rows{3}{\z}$,
        \item exactly $1$ interval per row in $\Z(\x)$, and
        \item at most $2$ intervals per row in $\rows{2}{\x} \setminus \left( \Z(\x) \cup \rows{3}{\z} \right)$.
    \end{enumerate}
\end{observation}

Furthermore, we define two variables with respect to the sets $\Z(\x)$ and $\rows{3}{\z}$.
Similarly to $\alp(i)$ being a fraction of the $\rows{1}{\x}$ rows, we define $\bet $ and $\gam $ as fractions of the set~$\rows{2}{\x}$ with respect to interval $\z$.

\begin{definition}
    \label{def:gamma_delta}
    Given $\z \in \rows{2}{\x}$, 
    \begin{enumerate}[(a)]
        \item $\bet \in [0,1]$ denotes the fraction of rows $\mathcal{R} \in \rows{2}{\x}$ where the interval $I^\prime$ in $\mathcal{R} \cap \nb(\x)$ is in $\zbar$, and
        \item $\gam \in [0,1]$ denotes the fraction of rows $\mathcal{R} \in \rows{2}{\x}$ where the interval $I^\prime$ in $\mathcal{R} \cap \nb(\x)$ is in $\rows{3}{\z}$.
    \end{enumerate}
\end{definition}

Now we are equipped to compute an upper bound on the color of interval $\z$.

\begin{lemma}
    \fullversion{\label{lem:any_z}}
    The color of interval $\z$ is at most $c(\z) \leq \omega + \alp(\z) \cdot \nrows{1}{\x} - \bet  \cdot \nrows{2}{\x} + \gam  \cdot \nrows{2}{\x}  + \nrows{2}{\x} + \nrows{3}{\x}$. 
\end{lemma}

\begin{proof}
    Using the Neighborhood bound and Observations~\ref{obs:any_r2} and~\ref{obs:any_z}, we can describe the maximum color of interval $\z$ as $1$ plus the number of intersections per type of row.
    Observe that by doing this, we count the row that contains interval $\z$ itself, and since $\z$ cannot intersect any interval in this row, we may subtract at least $1$.
    Then it follows that
    \begin{align*}
            \ffc(\z) &\leq \nrows{0}{\x} + (1 + \alp(\z)) \cdot \nrows{1}{\x} + \bet \cdot \nrows{2}{\x} + \gam \cdot 3\nrows{2}{\x} \\&\hspace{2cm}+ (1 - \bet - \gam) \cdot 2\nrows{2}{\x} + 2\nrows{3}{\x} + 1 - 1\\
            &= \omega + \alp(\z) \cdot \nrows{1}{\x} - \bet  \cdot \nrows{2}{\x} + \gam \cdot \nrows{2}{\x} + \nrows{2}{\x} + \nrows{3}{\x}
        \end{align*}
\end{proof}


Next, we make a case distinction based on the relative sizes of the sets $\Z(\x)$ and $\rows{3}{\z}$.
\paragraph*{Case (2.a): $|\Z(\x)| \geq \nrows{3}{\z}$}
We start with the case where the number of intervals in the set $\Z(\x)$ is at least as large as the number of intervals in the set $\rows{3}{\z}$.

\medskip 

\runtitle{Bound the size of $\pivotset$.}

\begin{lemma}
    \fullversion{\label{lem:pivotset_z}}
    When $\ffc(\z) > \ffc(\y)$, by selecting $\z$ as pivot $\pivot$, $ |\pivotset| = (1 - \alp(\z)) \cdot \nrows{1}{\x}$.
\end{lemma}

\begin{proof}
    First, by the definition of $\rows{0}{\x}$, $\x$ does not intersect any interval in $\rows{0}{\x}$.
    Thus, no interval in $\rows{0}{\x}$ can be in $\pivotset$.
    Next, all intervals in $\rows{2}{\x}$ are either in $\lmr(\x)$ or $\nlmr(\x)$.
    Since $\ffc(\z) > \ffc(\y)$, it follows that all intervals in $\rows{2}{\x}$ have a color below $\ffc(\z)$ and cannot contribute to $\pivotset$.
    Similarly, as all intervals in $\rows{3}{\x}$ are in $\lmr(\x)$, all intervals in $\rows{3}{\x}$ have a color below $\ffc(\z)$ and cannot contribute to $\pivotset$.
    The remaining intervals are those in $\rows{1}{\x}$.
    By definition, $\alp(\x) \cdot \nrows{1}{\x}$ of these intervals have a color at most $\ffc(\z)$.
    Hence, the remaining $(1 - \alp(\z) ) \cdot \nrows{1}{\x}$ are the only intervals intersecting $\x$ that have a color greater than $\ffc(\z)$. That is, $|\pivotset| = (1 - \alp(\z)) \cdot \nrows{1}{\x}$.
\end{proof}

\medskip

\runtitle{Bound the color of $\ffc(\x)$.}

\fullversion{\begin{lemma}
    \fullversion{\label{lem:gamma<delta}}
    If $|\Z(\x)| \geq \nrows{3}{\z}$, then $\ffc(\x) \leq 2\omega$.
\end{lemma}}

\begin{proof}
    Assume that $\ffc(\z) > \ffc(\y)$ and interval $\x$ is a closed interval, as otherwise by Lemma~\ref{lem:c_u>=c_z} and Lemma~\ref{lem:open_interval_small_intersection} it follows that $\ffc(\x) \leq 2\omega$.
    From the definition of $\bet $ and $\gam $ it follows that since $|\Z(\x)| \geq \nrows{3}{\z}$, also $\bet  \geq \gam $.
    
    Now we can compute a bound on the color of interval $\x$, using the Pivot bound where we take interval $\z$ as $\pivot$.
    By Lemma~\ref{lem:any_z}, the color of interval $\z$ is bounded by $\ffc(\z) \leq \omega + \alp(\z) \cdot \nrows{1}{\x} - \bet  \cdot \nrows{2}{\x} + \gam  \cdot \nrows{2}{\x} + \nrows{2}{\x} + \nrows{3}{\x}$.
    And, by Lemma~\ref{lem:pivotset_z}, the size of $\pivotset$ equals $(1 - \alp(\z)) \cdot \nrows{1}{\x}$.
    Then it follows that,

    \begin{align*}
        \ffc(\x) &\leq \ffc(\pivot) + |\pivotset| + 1\\
        &= \ffc(\z) + (1 - \alp(\z)) \cdot \nrows{1}{\x} + 1 \\
        &\leq \omega + \nrows{1}{\x} - \bet  \cdot \nrows{2}{\x} + \gam  \cdot \nrows{2}{\x} + \nrows{2}{\x} + \nrows{3}{\x} + 1 \\
        &\leq \omega + \nrows{1}{\x} + \nrows{2}{\x} + \nrows{3}{\x} + 1 \\
        &\leq 2\omega
    \end{align*}
\end{proof}
\paragraph*{Case (2.b): $|\Z(\x)| < \nrows{3}{\z}$}
We continue with the case where the number of intervals in the set $\Z(\x)$ is strictly smaller than the number of intervals in the set $\rows{3}{\z}$.
This relation between $\Z(\x)$ and $\rows{3}{\z}$ indicates that the average number of intersections interval $\z$ has per row can be greater than $2$.
Then, it might be interesting to explore other options as a pivot.
For this we make another case distinction on the relation between the colors of interval $\y$ and interval $\zbar$.

\paragraph*{Case (2.b.i): $\ffc(\zbar) \geq \ffc(\y)$}
First, let us explore the case where the color of interval $\zbar$ is at least as large as the color of interval $\y$.

\medskip

\runtitle{Pick a pivot $\pivot$}
The condition that $\ffc(\zbar) \geq \ffc(\y)$ guarantees that all intervals in the set $\lmr(\x)$ have colors at most $\ffc(\zbar)$.
Therefore, for the sake of the size of the size of the set $\pivotset$, it is more efficient to pick interval $\zbar$ as a pivot compared to picking interval $\y$.
Therefore, we pick interval $\zbar$ as pivot $\pivot$.

\medskip

\runtitle{Bound the color $\ffc(\pivot)$}

As we now want to compute an upper bound on the color of interval $\zbar$, it is of importance to know where interval $\zbar$ might be located.

\begin{lemma}
    \label{lem:zbar_not_in_LMR}
    For any interval $I^\prime \in \rows{3}{\z}$, $I^\prime \not \in \lmr(\x)$
\end{lemma}

\begin{proof}
    Assume aiming towards a contradiction that $I^\prime \in \lmr(\x)$.
    \medskip

    \runtitle{(1) $I^\prime$ is a closed interval.} 
    Assume without loss of generality that interval $I^\prime$ intersects interval $\x$ on the left-hand side of interval $\x$.
    Since $I^\prime \in \rows{3}{\z}$, and $I^\prime$ is a closed interval, it follows that there are precisely two positions where interval $\z$ could lie.
    The first possible position is to the left of interval $I^\prime$.
    In this case, interval $\z$ and interval $\x$ cannot intersect.
    Then it follows that $\z \not \in \nlmr(\x)$, which is a contradiction by the definition of $\z$.
    The other possible position is to the right of interval $I^\prime$.
    In this case, interval $\z$ must be identical to interval $\x$, i.e., $\z \in \twins(\x)$.
    Then it follows that $\z \not \in \nlmr(\x)$, which is a contradiction by the definition of $\z$.
    
    \medskip
    
    \runtitle{(2) $I^\prime$ is an open interval.} 
    Since $I^\prime \in \lmr(\x)$ and $I^\prime$ is an open interval, it follows that $I^\prime \in \twins(\x)$.
    Similarly, since $I^\prime \in \rows{3}{\z}$, and $I^\prime$ is an open interval, it follows that $I^\prime \in \twins(\z)$.
    An interval can only be in both $\twins(\x)$ and $\twins(\z)$ if interval $\x$ and interval $\z$ are identical, i.e., $\z \in \twins(\x)$.
    Then it follows that $\z \not \in \nlmr(\x)$, which is a contradiction by the definition of $\z$.
\end{proof}

\begin{figure}
\centering
\begin{minipage}{0.49\textwidth}
\centering
\includegraphics[width=\textwidth]{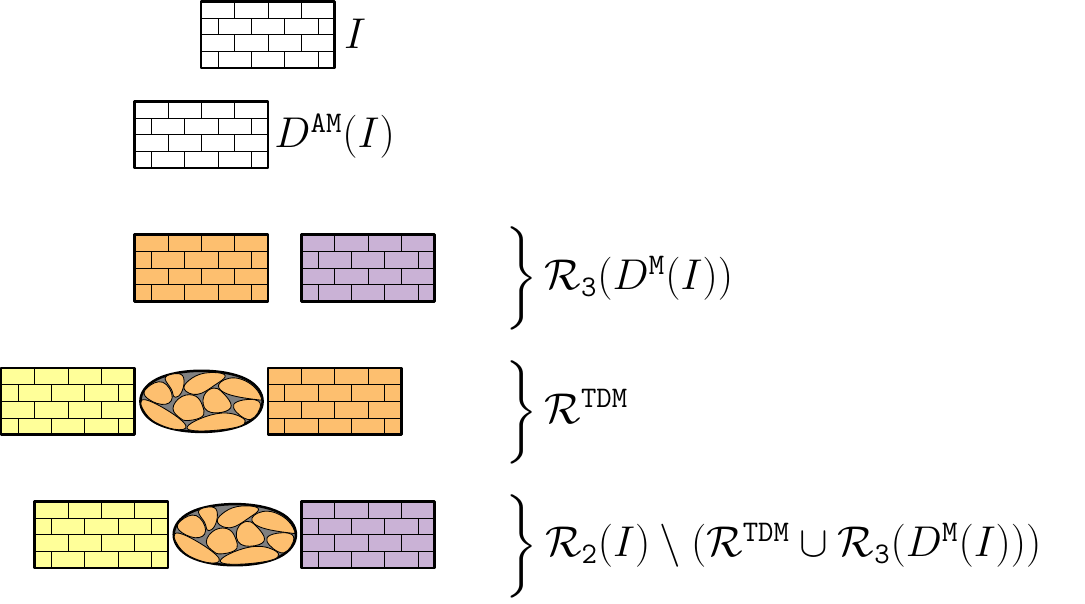}
\end{minipage}%
\hspace{0.01\textwidth}%
\begin{minipage}{0.49\textwidth}
\centering
\includegraphics[width=\textwidth]{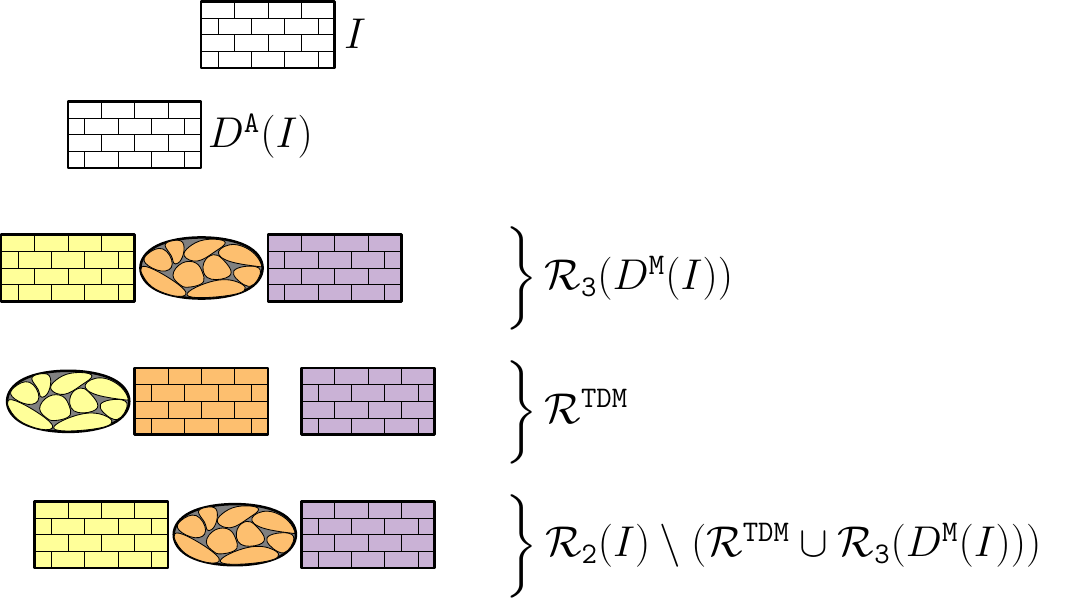}
\end{minipage}%
\caption{
An exemplary non-exhaustive list of possible rows per type of $\rows{2}{\x}$.
Yellow intervals intersect only interval $\zbar$ (resp. $\y$), purple intervals intersect only $\x$ and orange intervals intersect both $\zbar$ (resp. $\y$) and $\x$.}
\label{fig:y_intersections}
\end{figure}

Note that Observation~\ref{obs:any_r2} remains true for interval $\zbar$.
Now, let us make an additional observation regarding the intersections interval $\zbar$ has with intervals in $\rows{2}{\x}$ rows (also see Figure~\ref{fig:y_intersections}):

\begin{observation}
    \label{obs:any_z_bar}
    Interval $\zbar$ intersects
    \begin{enumerate}[(a)]
        \item exactly $1$ interval per row contained in $\rows{3}{\z}$,
        \item exactly $3$ intervals per row contained in $\Z(\x)$, and
        \item at most $2$ intervals per row contained in $\rows{2}{\x} \setminus (\Z(\x) \cup \rows{3}{\z} )$.
    \end{enumerate}
\end{observation}

Furthermore, it is important to specify how many of the intervals in $\rows{2}{\x} \setminus \rows{3}{\z}$ are assigned a color larger than $\ffc(\zbar)$ or smaller than $\ffc(\zbar)$.
This was previously not necessary, when we considered interval $\z$, which is, by definition, the interval in $\rows{2}{\x}$ assigned the largest color.
Analogously, we do not need to consider the intervals in $\rows{3}{\z}$ as per definition interval $\zbar$ is the largest colored interval in this set.

\begin{definition}
    \label{def:delta}
    Given any interval $\hat{I}$, $\delt(\hat{I}) \in [0, 1]$ denotes the fraction of rows $\mathcal{R} \in \rows{2}{\x} \setminus \rows{3}{\z}$ where interval $I^\prime$ in $\mathcal{R} \cap \nb(\x) \cap \nb(\z)$ satisfies $\ffc(I^\prime) \leq \ffc(\hat{I})$.
\end{definition}

Now we are equipped to compute a bound on the color of interval $\zbar$.

\begin{lemma}
    \label{lem:any_z_bar}
    The color of interval $\zbar$, $\ffc(\zbar) \leq \nrows{0}{\x} + (1 + \alp(\zbar)) \cdot \nrows{1}{\x} + \delt(\zbar) \cdot (1 - \gam ) \cdot 2\nrows{2}{\x} + \bet  \cdot \nrows{2}{\x} + \gam  \cdot \nrows{2}{\x} + 2\nrows{3}{\x}$.
\end{lemma}

\begin{proof}
    By the Neighborhood bound $\ffc(\zbar) \leq 1 + |\mathcal{N}(\zbar)|$.
    By Observation~\ref{obs:any_r2}, there are at most $\nrows{0}{\x} + 2\nrows{1}{\x} + 2\nrows{3}{\x}$ intervals in $\rows{0}{\x}, \rows{1}{\x}$ and $\rows{3}{\x}$ that are in $\mathcal{N}(\zbar)$.
    However, by Definition~\ref{def:alpha}, $(1 - \alp(\zbar))$ of those intervals have a color larger than $\ffc(\zbar)$ and hence we do not consider them for this bound.
    By Observation~\ref{obs:any_z_bar}, there are at most $\nrows{3}{\z} + 3\cdot|\Z(\x)| + 2\cdot|\rows{2}{\x}\setminus(\Z(\x)\cup\rows{3}{\z})| = \gam \cdot \nrows{2}{\x} + \bet \cdot 3\nrows{2}{\x} + (1 - \bet - \gam) \cdot 2\nrows{2}{\x}$ intervals in $\rows{2}{\x}$ that are in $\mathcal{N}(\zbar)$.
    However, by Definition~\ref{def:delta}, $(1 - \delt(\zbar)) \cdot (\bet \cdot 2\nrows{2}{\x} + (1 - \bet - \gam) \cdot 2\nrows{2}{\x})$ of those intervals have a color larger than $\ffc(\zbar)$ and hence we do not consider them for this bound.
    Observe that by this bound, we count the row that contains interval $\zbar$ itself.
    Since $\zbar$ cannot intersect any interval on this row, we should subtract at least 1 from this bound.
    Thus,

    \begin{align*}
            \ffc(\zbar) &\leq \nrows{0}{\x} + 2\nrows{1}{\x} - (1 - \alp(\zbar)) \cdot \nrows{1}{\x} + 2\nrows{3}{\x}  
            \\ &\hspace{1cm} + \bet  \cdot 3\nrows{2}{\x} + \gam  \cdot \nrows{2}{\x} + (1 - \bet  - \gam ) \cdot 2\nrows{2}{\x}
            \\ &\hspace{1cm} - (1 - \delt(\zbar)) \cdot \left( \bet  \cdot 2\nrows{2}{\x} + (1 - \bet  - \gam ) \cdot 2\nrows{2}{\x} \right) -1 + 1\\
            &= \nrows{0}{\x} + (1 + \alp(\zbar)) \cdot \nrows{1}{\x} + \delt(\zbar) \cdot (1 - \gam ) \cdot 2\nrows{2}{\x} + \bet  \cdot \nrows{2}{\x} \\&\hspace{1cm} + \gam  \cdot \nrows{2}{\x} + 2\nrows{3}{\x}
        \end{align*}
\end{proof}

\medskip

\runtitle{Bound the size of $\pivotset$.}

\begin{lemma}
    \label{lem:pivotset_zbar}
    By selecting $\zbar$ as pivot $\pivot$, $|\pivotset| = (1-\alp(\zbar)) \cdot \nrows{1}{\x} + (1 - \delt(\zbar) \cdot (1 - \gam) \cdot 2\nrows{2}{\x}$.
\end{lemma}

\begin{proof}
    First, by the definition of $\rows{0}{\x}$, $\x$ does not intersect any interval in $\rows{0}{\x}$.
    Thus, no interval in $\rows{0}{\x}$ can be in $\pivotset$.
    Next, all intervals in $\rows{3}{\x}$ are in the set $\lmr(\x)$.
    Thus, all these intervals are assigned a color below $\ffc(\zbar)$ and cannot contribute to $\pivotset$.
    Of the intervals in $\rows{1}{\x}$, by definition of $\alp$, only $(1 - \alp(\zbar))\cdot \nrows{1}{\x}$ intervals have a color larger than $\ffc(\zbar)$, and can contribute to set $\pivotset$.
    Of the $\rows{2}{\x}$ rows, out of the intervals intersecting $\x$ that are in the rows that are also in the set $\rows{3}{\z}$, none have a color larger than $\ffc(\zbar)$ by the definition of $\zbar$.
    Then, of the remaining $(1 - \gam) \cdot \nrows{2}{\x}$ $\rows{2}{\x}$ rows, by the definition of $\delt$, only $(1 - \delt(\zbar)) \cdot (1 - \bet - \gam) \cdot 2\nrows{2}{\x}$ intervals have a color larger than $\ffc(\zbar)$ and can contribute to the set $\pivotset$.
    Then it follows that $|\pivotset| = (1 - \alp(\zbar))\cdot \nrows{1}{\x} + (1 - \delt(\zbar)) \cdot (1 - \gam) \cdot 2\nrows{2}{\x} $
\end{proof}

\medskip

\runtitle{Bound the color $\ffc(\x)$.}

Now we can use this result to prove an upper bound on the color of interval $\x$, when the color of interval $\zbar$ is at least as large as the color of interval $\y$.

\begin{lemma}
    \label{lem:cy>=cu}
    If $\ffc(\zbar) \geq \ffc(\y)$, then $\ffc(\x) \leq 2\omega$   
\end{lemma}

\begin{proof}
    By Lemma~\ref{lem:gamma<delta}, we assume that $|\Z(\x)| < \nrows{3}{\z}$, and thus $\bet  < \gam$.
    By Lemma~\ref{lem:any_z_bar}, $\ffc(\zbar) \leq \nrows{0}{\x} + (1 + \alp(\zbar)) \cdot \nrows{1}{\x} + \delt(\zbar) \cdot (1 - \gam) \cdot 2\nrows{2}{\x} + \bet  \cdot \nrows{2}{\x} + \gam  \cdot \nrows{2}{\x} + 2\nrows{3}{\x}$.
    By Lemma~\ref{lem:pivotset_zbar}, $|\pivotset| = (1-\alp(\zbar)) \cdot \nrows{1}{\x} + (1 - \delt(\zbar) \cdot (1 - \gam) \cdot 2\nrows{2}{\x}$.
    It follows that,
    
    \begin{align*}
        \ffc(\x) &\leq \ffc(\pivot) + |\pivotset| + 1\\
        &\leq \ffc(\zbar) + (1 - \alp(\zbar)) \cdot \nrows{1}{\x} + ( 1 -  \delt(\zbar) ) \cdot (1 - \gam ) \cdot 2\nrows{2}{\x} + 1\\
        &\leq \nrows{0}{\x} + 2\nrows{1}{\x}  + \bet  \cdot \nrows{2}{\x} + \gam  \cdot \nrows{2}{\x} + (1 - \gam ) \cdot 2\nrows{2}{\x} + 2\nrows{3}{\x} + 1\\
        &< \nrows{0}{\x} + 2\nrows{1}{\x}  + \bet  \cdot \nrows{2}{\x} + (1 - \bet ) \cdot \nrows{2}{\x} + \nrows{2}{\x} + 2\nrows{3}{\x} + 1\\
        &\leq 2\omega
    \end{align*}
\end{proof}

\paragraph*{Case (2.b.ii): $\ffc(\y) > \ffc(\zbar)$ }

In order to prove Theorem~\ref{thm:any} we still need to consider the tough kernel of the analysis.
That is, the case where $\ffc(\zbar) < \ffc(\y) < \ffc(\z)$ and $\bet  < \gam $.
For this case we abandon our usual mechanism slightly, where instead of focusing on a single pivot $\pivot$ and a single set $\pivotset$, we explore the option where two distinct pairs of a pivot and a set cannot both admit a large solution.

\medskip

\runtitle{Pick the pivots $\pivot$.}
Although we have reasoned earlier that picking interval $\y$ and interval $\z$ as the pivot is not sufficient to have an upper bound on $\ffc(\x)$ that is smaller than $3\omega$, we can deal with this tough case by expressing the bound obtained by using $\y$ or $\z$ as the pivot in terms of $|\Z(\x)|$.

\medskip

\runtitle{Bound the color of $\pivot$}
Let us first take a closer look at the color of interval $\y$.
Although Lemma~\ref{lem:integral_uvw} is technically still correct, knowing that $\ffc(\zbar) < \ffc(\y)$, we can be slightly more precise about which intersections are assigned a color greater than $\ffc(\y)$.
In order to be more precise, let us first make additional observations regarding the intersections of interval $\y$ with interval in $\rows{2}{\x}$ rows (also see Figure~\ref{fig:y_intersections}).

\begin{observation}
    \label{obs:any_y}
    Interval $\y$ intersects
    \begin{enumerate}[(a)]
        \item at most $1$ interval that is not in $\nb(\x)$ per row contained in $\rows{2}{\x}$, and
        \item at most $1$ interval that is in $\nb(\x)$ per row contained in $\rows{3}{\z}$,
        \item at most $1$ intervals that is in $\nb(\x)$ per row contained in $\Z(\x)$, and
        \item at most $1$ intervals that is in $\nb(\x)$ per row contained in $\rows{2}{\x} \setminus (\Z(\x) \cup \rows{3}{\z} )$.
    \end{enumerate}
\end{observation}

\begin{lemma}
    \label{lem:any_uvw}
    The color of interval $\y$,
    $\ffc(\y) \leq 2\nrows{0}{\x} + 2\nrows{1}{\x} + \alp(\y) \cdot \nrows{1}{\x} + \gam  \cdot \nrows{2}{\x} + \nrows{2}{\x} + \delt(\y) \cdot (1 - \gam ) \cdot \nrows{2}{\x} + \nrows{3}{\x}$. 
\end{lemma}

\begin{proof}
    By the Neighborhood bound, $\ffc(\y) \leq 1 + |\mathcal{N}(\y)|$.
    By Observation~\ref{obs:integral}, there are at most $ 2\nrows{0}{\x} + 2\nrows{1}{\x} + \nrows{3}{\x}$ intervals in $\rows{0}{\x}, \rows{1}{\x}$ and $\rows{3}{\x}$ that are in $\nb(\y)$.
    Together with the at most $\alp(\y) \cdot \nrows{1}{\x}$ intervals that were assigned a color below $\ffc(\y)$ which are either in $\twins(\x)$ or do not intersect $\y$ and are in $\nlmr(\x)$.
    
    By Observation~\ref{obs:any_y}, there are at most $\nrows{2}{\x} + \gam \cdot \nrows{2}{\x} + \bet \cdot \nrows{2}{\x} + (1 - \gam - \bet) \cdot \nrows{2}{\x}$ intervals that are in $\rows{2}{\x}$ and in $\nb(\y)$.
    By our assumption that $\ffc(\zbar) < \ffc(\y)$, the $\gam  \cdot \nrows{2}{\x}$ intervals in the set $\rows{3}{\z} \cap \nb(\x)$, must have a color smaller than $\ffc(\y)$.
    For the remaining $(1 - \gam) \cdot \nrows{2}{\x}$ intervals that are in $\nb(\x)$, by the definition of $\delt$,  $(1 - \delt(\y)) \cdot (1 - \gam) \cdot \nrows{2}{\x}$ are assigned a color larger than $\ffc(\z)$ and hence we do not consider them for this bound.
    
    Observe that by this bound we count the row that contains interval $\y$ itself.
    Since interval $\y$ cannot intersect any interval on this row, we should subtract at least 1 from this bound. 
    Thus,

    \begin{align*}
            \ffc(\y) &\leq |\nb(\y)| + 1\\
            &\leq 2\nrows{0}{\x} + 2\nrows{1}{\x} + \alp(\y) \cdot \nrows{1}{\x} + \gam  \cdot \nrows{2}{\x} \\&\hspace{1cm} + \nrows{2}{\x} + \delt(\y) \cdot (1 - \gam ) \cdot \nrows{2}{\x} + \nrows{3}{\x}
        \end{align*}    
\end{proof}

As for the bound on the color of interval $\z$, Lemma~\ref{lem:any_z} still suffices.

\medskip

\runtitle{Bound the sizes of $\pivotset$.}
Given that we use $\y$ as pivot $\pivot$, we bound the size of $\pivotset$.

\begin{lemma}
    \label{lem:pivotset_y_exact}
    When $\ffc(\y) \geq \ffc(\zbar)$, by selecting $\y$ as pivot $\pivot$, $|\pivotset| \leq (1 - \alp(\y)) \cdot \nrows{1}{\x} + (1 - \delt(\z)) \cdot (1 - \gam ) \cdot \nrows{2}{\x} + (1 - \gam ) \cdot \nrows{2}{\x}$
\end{lemma}

\begin{proof}
    First, by the definition of $\rows{0}{\x}$, $\x$ does not intersect any interval in $\rows{0}{\x}$.
    Thus, no interval in $\rows{0}{\x}$ can be in $\pivotset$.
    By the definition of $\alp$, only $(1 - \alp(\y) ) \cdot \nrows{1}{\x}$ intervals in $\rows{1}{\x}$ are assigned a color strictly larger than $\ffc(\y)$.
    Thus the only intervals in an $\rows{1}{\x}$ that are eligible for $\pivotset$ are those $(1 - \alp(\y) ) \cdot \nrows{1}{\x}$ intervals.
    Since $\ffc(\y) > \ffc(\zbar)$, no interval in an $\rows{3}{\z}$ row is assigned a color larger than $\ffc(\y)$.
    Thus, none of the $\gam \cdot 2\nrows{2}{\x}$ intervals in $\rows{3}{\z} \cap \nb(\x)$ can contribute to $\pivotset$.
    By the definition of $\delt$, only $(1 - \delt(\y) ) \cdot (1 - \gam) \cdot \nrows{2}{\x}$ intervals both in $\nb(\y)$ and $\rows{2}{\x}$ are assigned a color strictly larger than $\ffc(\y)$.
    Furthermore, all $(1 - \gam) \cdot \nrows{2}{\x}$ intervals which are in $\rows{2}{\x}$ but not in $\nb(\y)$ and not in $\rows{3}{\z}$ could potentially be assigned a color larger than $\ffc(\y)$ and therefore contribute to $\pivotset$.
    Finally, as all intervals in $\rows{3}{\x}$ are in $\lmr(\x)$, all intervals in $\rows{3}{\x}$ have a color below $\ffc(\y)$ and cannot contribute to $\pivotset$.
    Thus, $|\pivotset| \leq (1 - \alp(\y)) \cdot \nrows{1}{\x} + (1 - \delt(\z)) \cdot (1 - \gam ) \cdot \nrows{2}{\x} + (1 - \gam ) \cdot \nrows{2}{\x}$.
\end{proof}

As for the set $\pivotset$ when we pick interval $\z$ as pivot, Lemma~\ref{lem:pivotset_z} still suffices.

\medskip

\runtitle{Bound the size of $\ffc(\x)$.}

As the bound we are going to prove in the proof of Theorem~\ref{thm:any} is based on the size of set $\Z(\x)$, which in turn is strictly smaller than the size of set $\rows{2}{\x}$, an upper bound on $\nrows{2}{\x}$ improves the bound on the color of $\x$ we would otherwise obtain.
This upper bound on the number of rows in $\rows{2}{\x}$ is obtained by a lower bound on the number of rows in $\rows{3}{\x}$.

\begin{lemma}
    \label{lem:r3_lb}
    For any interval $\x \in \orderedInput$, if $\nrows{3}{\x} < \frac{1}{3}\omega$, then $\ffc(\x) \leq \frac{7}{3} \omega - 2$.
\end{lemma}

\begin{proof}
    We show this using the Neighborhood bound.
    Observe that $\nrows{0}{\x} \geq 1$, since interval $\x$ does not intersect any intervals on its own row.
    Furthermore, note that the Neighborhood bound is maximized when we maximize the number of intervals that intersect interval $\x$.
    That is, we maximize the bound when we maximize $\nrows{3}{\x}$.
    Then, $\nrows{3}{\x} = \frac{1}{3}\omega - 1$.
    For the remaining rows, we maximize the bound by maximizing $\nrows{2}{\x}$, and hence $\nrows{2}{\x} = \frac{2}{3}\omega$.
    Then, $\nrows{1}{\x} = 0$, and we get the following upper bound on the color of interval $\x$,
    \begin{align*}
        \ffc(\x) &\leq \nrows{1}{\x} + 2\nrows{2}{\x} + 3\nrows{3}{\x} + 1 \\
            &\leq 0 + 2 \cdot \frac{2}{3} \omega + 3 \cdot (\frac{1}{3}\omega - 1) + 1\\
            &= \frac{7}{3} \omega - 2
    \end{align*}
\end{proof}

Now we have shown that $\nrows{3}{\x} \geq \frac{1}{3}\omega$, it simply follows that $\nrows{2}{\x} < \frac{2}{3}\omega$.

\begin{corollary}
    \label{cor:r2_bound}
    For any interval $\x \in \orderedInput$, if $\nrows{2}{\x} \geq \frac{2}{3} \omega$, then $\ffc(\x) \leq \frac{7}{3} \omega - 2$. 
\end{corollary}

We have now shown everything we need to show in order to compute a bound on the color of interval $\x$.

\paragraph*{Proof of Theorem~\ref{thm:any}.}
\emph{For any interval $\x \in \orderedInput$, the color of $\x$ is at most $\ffc(\x) \leq \result$.}

\begin{proof}
    By Lemma~\ref{lem:open_interval_small_intersection},~\ref{lem:integral_open_and_closed_assumtion_not_hold}, \ref{lem:c_u>=c_z}, \ref{lem:gamma<delta} and \ref{lem:cy>=cu}, $\ffc(\x) \leq 2\omega$ if $\x$ is open, $\nrows{1}{\x} \leq \nrows{3}{\x} - 2$, $\ffc(\y) < \ffc(\z)$, $\bet  < \gam $ or $\ffc(\zbar) < \ffc(\y)$.
    Thus in the following we focus on closed intervals with $\nrows{1}{\x} \leq \nrows{3}{\x} - 2$, $\ffc(\zbar) < \ffc(\y) < \ffc(\z)$ and $\bet  < \gam $.

    Now we can compute the first bound on the color of interval $\x$, using the Pivot bound with interval $\y$ as pivot $\pivot$.
    According to Lemma~\ref{lem:any_uvw}, $\ffc(\y) \leq  2\nrows{0}{\x} + 2\nrows{1}{\x} + \alp(\y) \cdot \nrows{1}{\x} + \gam  \cdot \nrows{2}{\x} + \nrows{2}{\x} + \delt(\z) \cdot (1 - \gam ) \cdot \nrows{2}{\x} + \nrows{3}{\x}$ and, by taking $\y$ as pivot $\pivot$,  Lemma~\ref{lem:pivotset_y_exact}, $|\pivotset| \leq (1 - \alp(\y)) \cdot \nrows{1}{\x} + (1 - \delt(\z)) \cdot (1 - \gam ) \cdot \nrows{2}{\x} + (1 - \gam ) \cdot \nrows{2}{\x}$.
    Thus,

    \begin{align*}
        \ffc(\x) &\leq \ffc(\pivot) + |\pivotset| + 1\\
        &\leq \ffc(\y) + (1 - \alp(\y)) \cdot \nrows{1}{\x} + (1 - \delt(\z)) \cdot (1 - \gam ) \cdot \nrows{2}{\x} \\ &\hspace{1cm} + (1 - \gam ) \cdot \nrows{2}{\x} + 1 \\
        &\leq 2\nrows{0}{\x} + 3\nrows{1}{\x} + 2\nrows{2}{\x} + (1 - \gam ) \cdot \nrows{2}{\x} + \nrows{3}{\x} + 1 \\
        &\leq 2\nrows{0}{\x} + 2\nrows{1}{\x} + (1 - \gam ) \cdot \nrows{2}{\x} + 2\nrows{2}{\x} + 2\nrows{3}{\x} - 1\\
        &= 2\omega + (1 - \gam ) \cdot \nrows{2}{\x} - 1
    \end{align*}

    Next we move to the other bound.
    By taking $\z$ as pivot $\pivot$, according to Lemma~\ref{lem:any_z}, $\ffc(\z) \leq \omega + \alp(\z) \cdot \nrows{1}{\x} - \bet  \cdot \nrows{2}{\x} + \gam  \cdot \nrows{2}{\x} + \nrows{2}{\x} + \nrows{3}{\x}$, and Lemma~\ref{lem:pivotset_z}, $|\pivotset| = (1 - \alp(\z))\cdot\nrows{1}{\x}$.
    Then, it follows that,

    \begin{align*}
        \ffc(\x) &\leq \ffc(\pivot) + |\pivotset| + 1\\
        &\leq \ffc(\z) + (1 - \alp(\z)) \cdot \nrows{1}{\x} + 1 \\
        &\leq \omega + \nrows{1}{\x} - \bet  \cdot \nrows{2}{\x} + \gam  \cdot \nrows{2}{\x} + \nrows{2}{\x} + \nrows{3}{\x} + 1\\ 
        &\leq 2\omega + \gam  \cdot \nrows{2}{\x} - 1 
    \end{align*}

    We have proven that simultaneously, the color of interval $\x$ is at most $\ffc(\x) \leq 2\omega + \gam  \cdot \nrows{2}{\x} - 1$ and at most $\ffc(\x) \leq 2\omega + (1 - \gam ) \cdot \nrows{2}{\x} - 1$.
    Then, it follows that for any value of $\gam $ the color of interval $\x$ is at most $\ffc(\x) \leq 2\omega + \frac{1}{2}\cdot \nrows{2}{\x} - 1$
    Then, it follows from Corollary~\ref{cor:r2_bound} that the number of $\rows{2}{\x}$ rows must be less than $\nrows{2}{\x}  < \frac{2}{3}\omega$, as otherwise the color of interval $\x$ is bounded by $\ffc(\x) \leq \frac{7}{3} \omega - 2$.
    Thus, the color of interval $\x$ is at most $\ffc(\x) < 2\omega + \frac{1}{3}\cdot \omega - 1 = \frac{7}{3}\omega - 1$.
    Which, by the integrality of $\ffc(\x)$, is at most $\ffc(\x) \leq \result$  
\end{proof}
}

\section{Conclusion}
\label{section:conclusion}
In this work, we develop a sophisticated counting method based on the Pivot bound and show that \ff{} uses at most $2\omega$ colors in the case that all open and closed unit-length intervals have integral endpoints, which matches the lower bound.
We also show that \ff{} uses at most $\result$ colors when the input open or closed unit-length intervals have arbitrary endpoints.
It remains open to find a tight bound for this problem.

In an attempt to improve the current upper bound for the number of colors used by the \ff{} algorithm for the general case, it may be of interest to investigate the number of colors used by the \ff algorithm for the so-called bounded length intervals as introduced by Chybowska-Sok\'{o}l et al.~\cite{DBLP:journals/ejc/ChybowskaSokol24}.
The technique and the results on open/closed unit-length intervals developed in this work serve as a first step towards investigating \ff's performance via gradually relaxing the lengths bound.

\newpage



\bibliography{main}

\newpage















\end{document}